\documentclass[
    ,final            
  ,numberedheadings 
  ]
  {aipproc}

\layoutstyle{8x11single}

\begin{document}

\title{Mesoscopic effects in quantum dots, nanoparticles and nuclei}


\author{Y. Alhassid}{
  address={Center for Theoretical Physics, Sloane Physics Laboratory,
Yale University, New Haven, CT 06520, U.S.A.}
}

\begin{abstract}
 We discuss mesoscopic effects in quantum dots, nanoparticles and nuclei.  In quantum dots, we focus on the statistical regime of dots whose single-electron dynamics are chaotic. Random matrix theory methods, developed to explain the statistics of neutron resonances in compound nuclei, are useful in describing the mesoscopic fluctuations of the conductance in such dots.  However, correlation effects beyond the charging energy are important in almost-isolated dots. In particular, exchange and residual interactions are necessary to obtain a quantitative description of the mesoscopic fluctuations.  Pairing correlations are important in metallic nanoparticles and nuclei. Nanoparticles smaller than $\sim 3$ nm and nuclei are close to the fluctuation-dominated regime in which the Bardeen-Cooper-Schrieffer theory is not valid. Despite the large fluctuations, we find signatures of pairing correlations in the heat capacity of nuclei. These signatures depend on the particle-number parity of protons and neutrons.
\end{abstract}

\maketitle


\section{Introduction}

   Mesoscopic systems can include a large number of particles yet are sufficiently small  that finite-size quantum effects are important. A good example of a mesoscopic system is a quantum dot, a sub-micron scale conducting device in which up to several thousand electrons are confined by electrostatic potentials \cite{QD-rev}. Another example is a nanoparticle -- an ultrasmall metallic grain \cite{nano01}. The discreteness of the energy spectrum of a quantum dot or a nanoparticle becomes important at the low temperatures achieved in laboratory experiments.

 The transport properties (i.e., conductance) of quantum dots and nanoparticles can be measured by connecting them to leads. Linear conductance measurements provide information on the ground state of the system, while properties of excited states can be extracted from non-linear conductance experiments. Furthermore, by changing a gate voltage, it is possible to control the number of electrons in the device.

To a certain extent a compound nucleus can also be considered as a mesoscopic system. Several methods and ideas introduced in nuclear theory have indeed inspired developments in mesoscopic physics.

 In this article we review some of the mesoscopic effects in quantum dots, metallic nanoparticles and nuclei. We consider some analogies and differences between quantum dots and nuclei and between nanoparticles and nuclei.

  In small vertical dots, the confining potential is harmonic, leading to shell structure and the phenomenon of magic numbers familiar from atomic and nuclear physics \cite{QD-rev}.  The spectra of such dots can usually be understood in terms of shell effects plus an exchange interaction.

 For larger lateral dots, the symmetries responsible for the shell structure are broken and the single-electron dynamics in the dot become mostly chaotic \cite{alhassid00,aleiner02}.  In this regime, the main interest is in understanding the statistical properties of the spectrum and conductance sampled from dots with different shapes or magnetic fields rather than in explaining the precise properties of a particular dot.  In this regime random matrix theory (RMT) \cite{mehta,guhr98} has proven to be very useful. RMT was developed by Wigner, Dyson, Mehta and others in the 1950's and 1960's to explain the statistical properties of the neutron resonances in compound nuclei.

The simplest model of a quantum dot is the constant interaction (CI) model in which the electron-electron interaction is taken to be the classical charging energy.
In this limit the conductance is simply related to the single-particle wave functions in the dot, and the RMT approach is justified by the chaotic dynamics associated with the single-particle Hamiltonian.   However, in an almost-isolated dot, residual interactions beyond the charging energy are important and must be taken into account to achieve a quantitative understanding \cite{alhassid00,aleiner02}.  Furthermore, the spin degrees of freedom beyond the simple degeneracy factor become crucial.  The exchange interaction associated with the spin degrees of freedom has explained some of the discrepancies between the experiments and the predictions of the CI model.  Understanding the interplay between mean-field behavior and residual interactions is at the frontier of research in mesoscopic physics, making this field even more attractive to many-body theorists.

 Pairing correlations play an important role in metallic nanoparticles \cite{nano01}. Bulk metals become superconducting at low temperatures, a phenomenon explained by the Bardeen-Cooper-Schrieffer (BCS) theory \cite{BCS57}. BCS, as a mean field theory, is valid in the limit where the single-particle mean level spacing is small compared with the pairing gap.  However, in nanoparticles smaller than $\sim 3$ nm, this condition no longer holds and the grain is in a crossover between the BCS limit and the regime dominated by fluctuations of the order parameter.  The nucleus also belongs to this crossover regime since the single-nucleon mean level spacing within a shell is typically smaller than the pairing gap.  The effects of pairing correlations are much more subtle in this crossover regime \cite{nano01} and it is apriori not clear whether any thermal signatures of the bulk pairing transition survive in this regime. We expect to see some similarities between nanoparticles and nuclei in the manifestation of pairing correlations.

  We remark that in comparison with natural systems such as nuclei, quantum dots have the advantage that their parameters (e.g., size, shape, number of particles, coupling to the outside) can all be easily controlled by the experimentalist.  This makes quantum dots a practical tool to study finite-size effects in small quantum systems.

 Quantum dots are discussed in Sec.~\ref{quantum-dots}, where we focus on the statistical regime of almost-isolated ballistic dots. The CI model is briefly discussed in
Sec.~\ref{CI-model}. The application of RMT to the mesoscopic fluctuations of the conductance in the CI model is described in Sec.~\ref{RMT}. The CI plus RMT model can  explain only some of the data, and its limitations are summarized in
Sec.~\ref{limitations}. Sections \ref{universal-H} through \ref{residual} discuss correlation effects beyond the CI model. In Sec.~\ref{universal-H} we review the so-called universal Hamiltonian, obtained in the limit of a large number of electrons. In addition to the charging energy term, it contains an exchange interaction that depends on the total spin $\bf \hat S$ of the dot. In Sec.~\ref{conductance-int} we present a closed expression for the conductance in the presence of such an exchange interaction. This is a non-trivial many-body problem but through the use of spin projection it can be reduced to quantities that characterize non-interacting spinless electrons. The remaining residual interaction terms are small. However, they play an important role as is explained in
Sec.~\ref{residual}.

Nanoparticles are briefly discussed in Sec.~\ref{nanoparticles}, emphasizing the significance of pairing correlations in finite-size systems. Section \ref{nuclei} discusses pairing correlations in hot nuclei. In Sec.~\ref{SMMC} we explain how thermal properties of nuclei can be calculated beyond the mean field approximation using quantum Monte Carlo methods. In Sec.~\ref{pairing} we use these methods to study signatures of pairing correlations in the heat capacity of finite nuclei.  We find that these signatures depend on the parity of the proton and neutron numbers.

\section{Quantum dots}\label{quantum-dots}

\begin{figure}[h!]\label{dots}
  \includegraphics[height=.24\textheight]{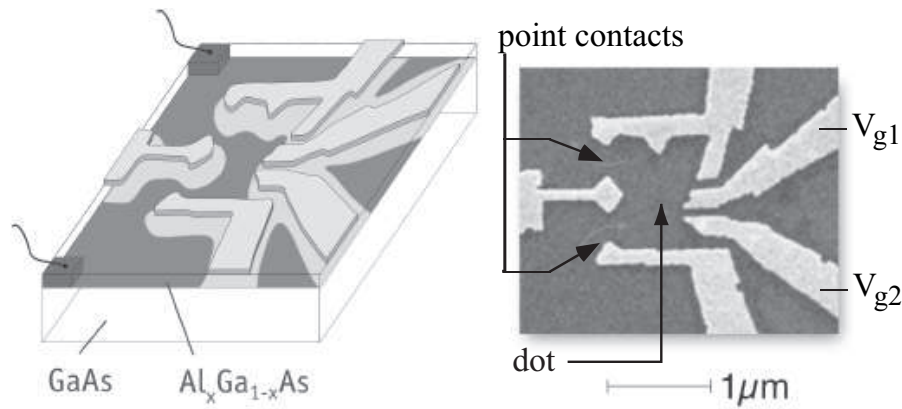}
  \caption{Quantum dot used in the experiment of Folk {\em et al.} \cite{folk96}. Right: a scanning electron micrograph of the dot (top view). Left: a schematic drawing.}
\end{figure}

  An example of a quantum dot \cite{folk96} is shown in Fig.~\ref{dots}. A quantum dot is typically made by forming a two-dimensional electron gas at the interface of a seminconductor heterostructure (e.g., GaAs/AlGaAs). Applying electrostatic potentials to metallic gates the electrons are confined to a 2D sub-micron region \cite{QD-rev}.

\vspace{3 mm}

  The transport properties (i.e., conductance) of a quantum dot are measured by connecting it to leads. At sufficiently low temperature (below $\sim 100$ mK), the coherence length $L_\phi$ of the electrons in the dot is larger than the size of the dot $L$ and the system is called mesoscopic. Phase decoherence of the electron results from inelastic scattering from other electrons or phonons.

 The electron in the dot scatters elastically from impurities with a mean free path $l$. The dot is diffusive when $l \ll L$. The dot is ballistic when it has little disorder, i.e., $l > L$. In the latter case, transport is dominated by scattering from the boundaries of the dot.

\begin{figure}[h!]\label{harmonic-dot}
  \includegraphics[height=.22\textheight]{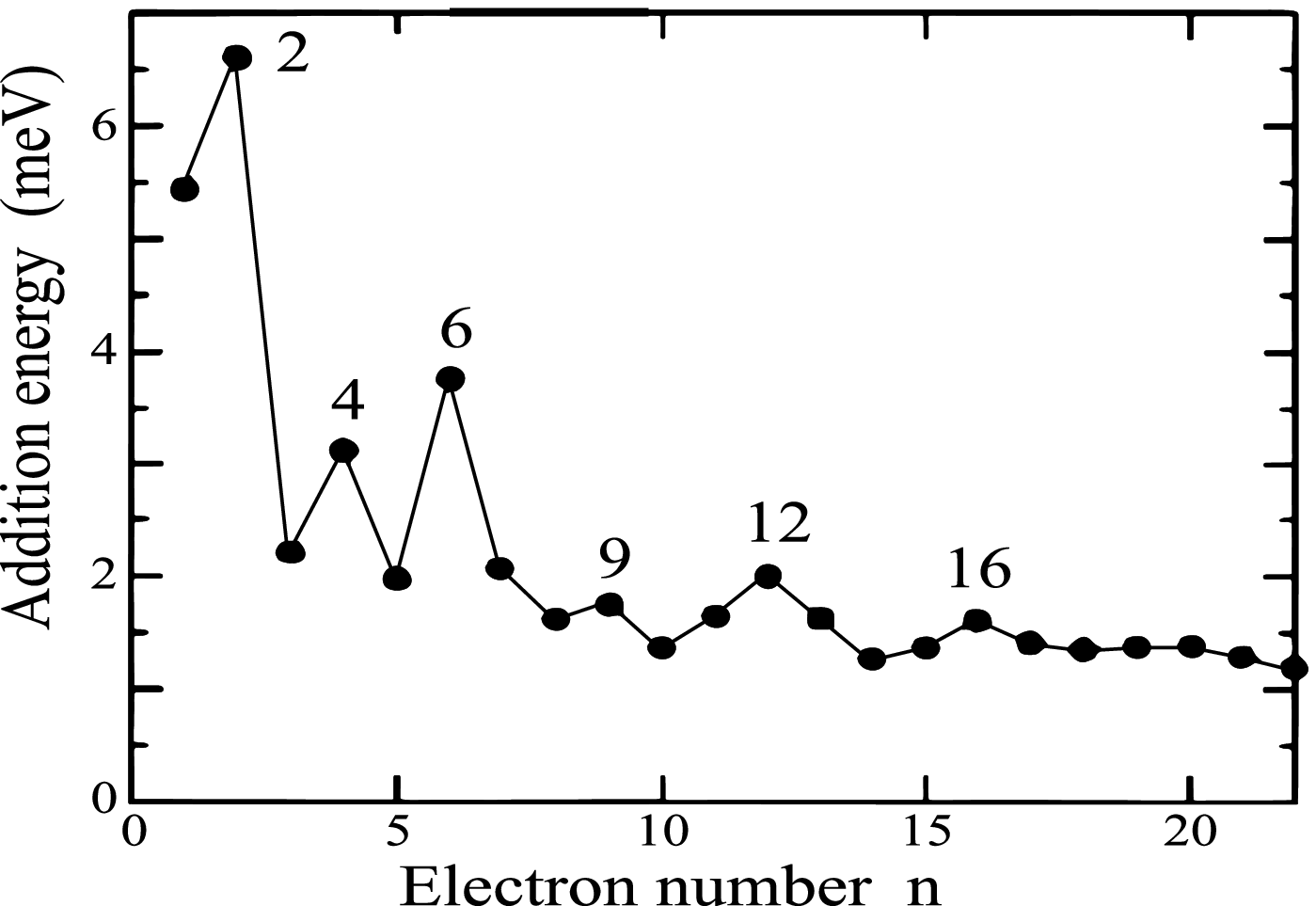}
  \caption{Shell structure observed in the addition energy of a small vertical dot. The dot has a shape of a disk with a harmonic-like confining potential. Maxima are observed for filled shells of a 2D spin-degenerate harmonic oscillator (magic numbers) and secondary maxima correspond to half-filled shells (following Hund's rule). From Ref.~\cite{tarucha96}.}
\end{figure}

  In small vertical dots (up to $N \sim 20$ electrons) the confining potential is often harmonic and leads to shell structure \cite{tarucha96}. The addition spectrum (i.e., the energy requires to add an electron into the dot) shown in Fig.~\ref{harmonic-dot} displays maxima at magic numbers $N=2, 6, 12$ that correspond to fermions with spin $1/2$ in a 2D harmonic potential.  Secondary maxima are observed at half-filled shells $N=4, 9, 16$. This is a result of Hund's rule from atomic physics -- the electrons are added with parallel spin up to mid-shell so as to maximize the (attractive) exchange interaction.

\vspace{3 mm}

\begin{figure}[h!]\label{ballistic-dot}
  \includegraphics[height=.15\textheight]{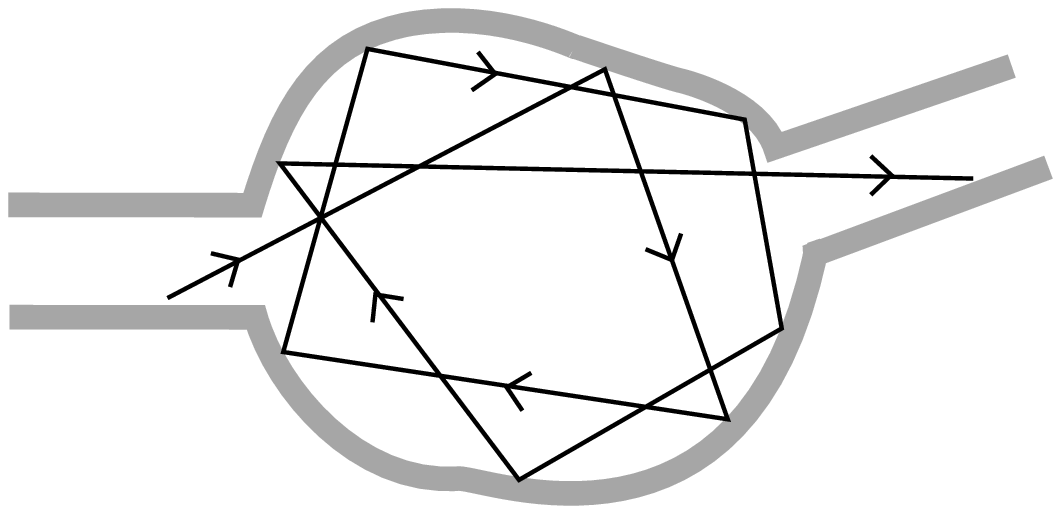}
  \caption{A schematic drawing of a ballistic quantum dot attached to two leads. A typical trajectory scattered several times from the boundaries of the dot before exiting.}
\end{figure}

  In larger dots the symmetries responsible for shell structure are broken and the single-particle dynamics are mostly chaotic (i.e., unstable with respect to the initial conditions)~\cite{gutzwiller90}.  Figure \ref{ballistic-dot} shows a schematic drawing of a ballistic dot with a chaotic trajectory scattered from the boundaries of the dot.  This regime is the statistical regime, in which we are interested in understanding the mesoscopic fluctuations of the conductance  \cite{alhassid00}. The relevant time scale in a ballistic dot is the time $\tau_{\rm f}$ to cross the dot. The statistical regime has some analogies with the compound nucleus, in which equilibration in $(n,\gamma)$ reactions is achieved through collisions of the incoming neutron with other nucleons (known as the ``doorway'' state mechanism). Here the relevant time scale is the equilibration time $\tau_{\rm eq}$.

\vspace{3 mm}

  The width of a resonance in a quantum dot is controlled by the coupling between the dot and the leads.  Denoting by $T_c$ the transmission coefficient for the electron to decay into channel $c$ in a lead and by $\tau_H$ the recurrence time (i.e., the Heisenberg time), the resonance width is given by
\begin{equation}
\Gamma= \hbar {1\over \tau_H} \sum_c T_c = {d \over 2 \pi} \sum T_c
\end{equation}
where we have used $\tau_H=h/ d$ ($d$ is the mean level spacing in the dot).

 An open dot is usually characterized by a large number of open channels with $T_c \sim 1$ so $\Gamma \gg d$, a situation analogous to the compound nucleus in the regime of overlapping resonances.  The conductance of an open dot exhibits Ericson fluctuations \cite{ericson60} versus a gate voltage.

  As we pinch off the point contacts, effective tunneling barriers are formed in the dot-lead interfaces and $T_c \ll 1$. In such almost-isolated dots, $\Gamma \ll \Delta$ and the conductance exhibits a series of peaks as a function of gate voltage.  An almost-isolated dot has some analogies with the regime of isolated neutron resonances in the compound nucleus. The top panels of Fig.~\ref{cn-dot} below compare the conductance peaks in a quantum dot with the neutron resonances observed in the total cross section to scatter thermal neutrons from heavy nuclei.

\subsection{Constant interaction (CI) model}\label{CI-model}

   The distance between the conductance peaks in a quantum dot is typically much larger than the single-particle mean level spacing $d$.  This is explained by a simple model, known as the constant interaction (CI) model. In the CI model, the Coulomb interaction energy is taken to be the classical electrostatic energy $e^2 N^2/2C$, where $N$ is the number of electrons in the dot and $C$ is the capacitance of the dot. The single-particle levels in the dot are spin-degenerate ($\sigma =\pm 1$) and depend only on the orbital label $\lambda$. The total Hamiltonian of the isolated dot in the CI model is
\begin{equation}\label{CI}
H_{\rm CI} = \sum_{\lambda \sigma} (\epsilon_\lambda - e \alpha V_g ) a^\dagger_{\lambda \sigma} a_{\lambda \sigma} + {e^2 \hat N^2 \over 2 C}\;.
\end{equation}
The term linear in the electron-number operator $\hat N = \sum_{\lambda \sigma} a^\dagger_{\lambda \sigma} a_{\lambda \sigma}$ describes the effect of the gate voltage $V_g$ ($\alpha$ is a constant determined by the ratio between the gate-dot capacitance and the total capacitance of the dot plus gate).

\subsubsection{Coulomb blockade}

\begin{figure}[h!]\label{CB}
  \includegraphics[height=.25\textheight]{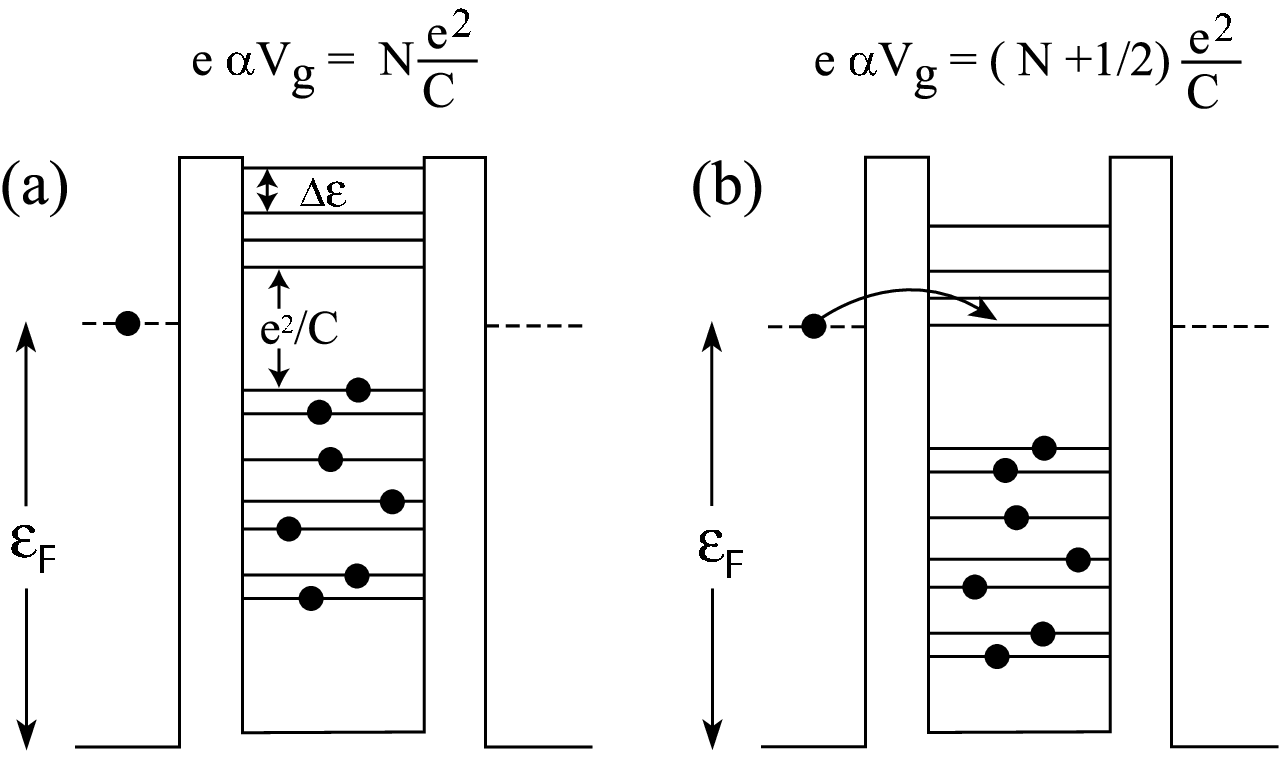}
  \caption{A schematic drawing of Coulomb blockade in a quantum dot: (a) when $e\alpha V_g= N e^2/C$ there is a charging energy gap on both sides of the Fermi energy that prevent the tunneling of an electron into the dot; (b) when $e \alpha V_g=(N+ 1/2)e^2/C$ there is no gap at the Fermi energy and an additional electron will tunnel into the dot.}
\end{figure}

  In general the tunneling of an additional electron into the dot is suppressed by the Coulomb repulsion of the electrons already in the dot, a phenomenon known as Coulomb blockade. By changing the gate voltage of the dot, it is possible to compensate for this Coulomb repulsion. The dot will conduct at its degeneracy points, i.e.  when $\epsilon_{\rm F} + E_{\rm gs} ( N-1) = E_{\rm gs}(N)$. Here $\epsilon_{\rm F}$ is the Fermi energy of the electron in the leads and $E_{\rm gs}(N)$ is the ground-state energy of a dot with $N$ electrons (we have assumed the temperature to be sufficiently low so that only the ground state of the dot participates in the conductance). Using Eq.~(\ref{CI}), we find that the effective Fermi energy $\tilde \epsilon_{\rm F} \equiv E_F + e \alpha V_g$ at the degeneracy points satisfies
\begin{equation}\label{peak-energies}
\tilde \epsilon_{\rm F}  = \left(N-\frac{1}{2}\right){e^2 \over  C} + \epsilon_N \;.
\end{equation}
The conductance displays a series of peaks at values of the gate voltage given by (\ref{peak-energies}). These are known as Coulomb-blockade peaks. The spacing $\Delta_2$ between two successive peaks is given by
\begin{equation}\label{peak-spacing}
\Delta_2 \equiv \Delta \tilde \epsilon_{\rm F} =  e^2/C + (\epsilon_{N + 1} - \epsilon_N) \;.
\end{equation}
The charging energy $e^2/C$ is typically much larger than the mean level spacing $d$ and therefore $\Delta_2 \approx e^2/C$ is approximately constant. The phenomenon of Coulomb blockade is demonstrated in Fig.~\ref{CB}.

\vspace{3 mm}

\subsubsection{Rate equations approach}

  When a typical width $\Gamma$ of a resonance in the dot much smaller than both the mean level spacing $d$ and the temperature (i.e. $\Gamma \ll d, kT$), the coherence between the dot and the leads can be ignored. It is then possible to use a rate equation approach to calculate the conductance through the dot \cite{beenakker91}. For low temperatures $kT \ll d$, only one resonance level $\lambda$ (near the Fermi energy) contributes to the conductance. The conductance peak is broadened by temperature and the important information is coded in the peak height $G$.  This peak height is given by $G \propto {e^2 \over h} {1 \bar \Gamma \over kT} g_\lambda$ where
\begin{equation}\label{g-CI}
 g_\lambda = {2 \over \Gamma} {\Gamma^l_\lambda \Gamma^r_\lambda \over  \Gamma^l_\lambda + \Gamma^r_\lambda} \;.
\end{equation}
Here $\Gamma^{l,(r)}_\lambda$ is the partial width of the level $\lambda$ to decay into the left (right) lead.

\subsection{Conductance fluctuations and random matrix theory}\label{RMT}

\begin{figure}[h!]\label{cn-dot}
  \includegraphics[height=.35\textheight]{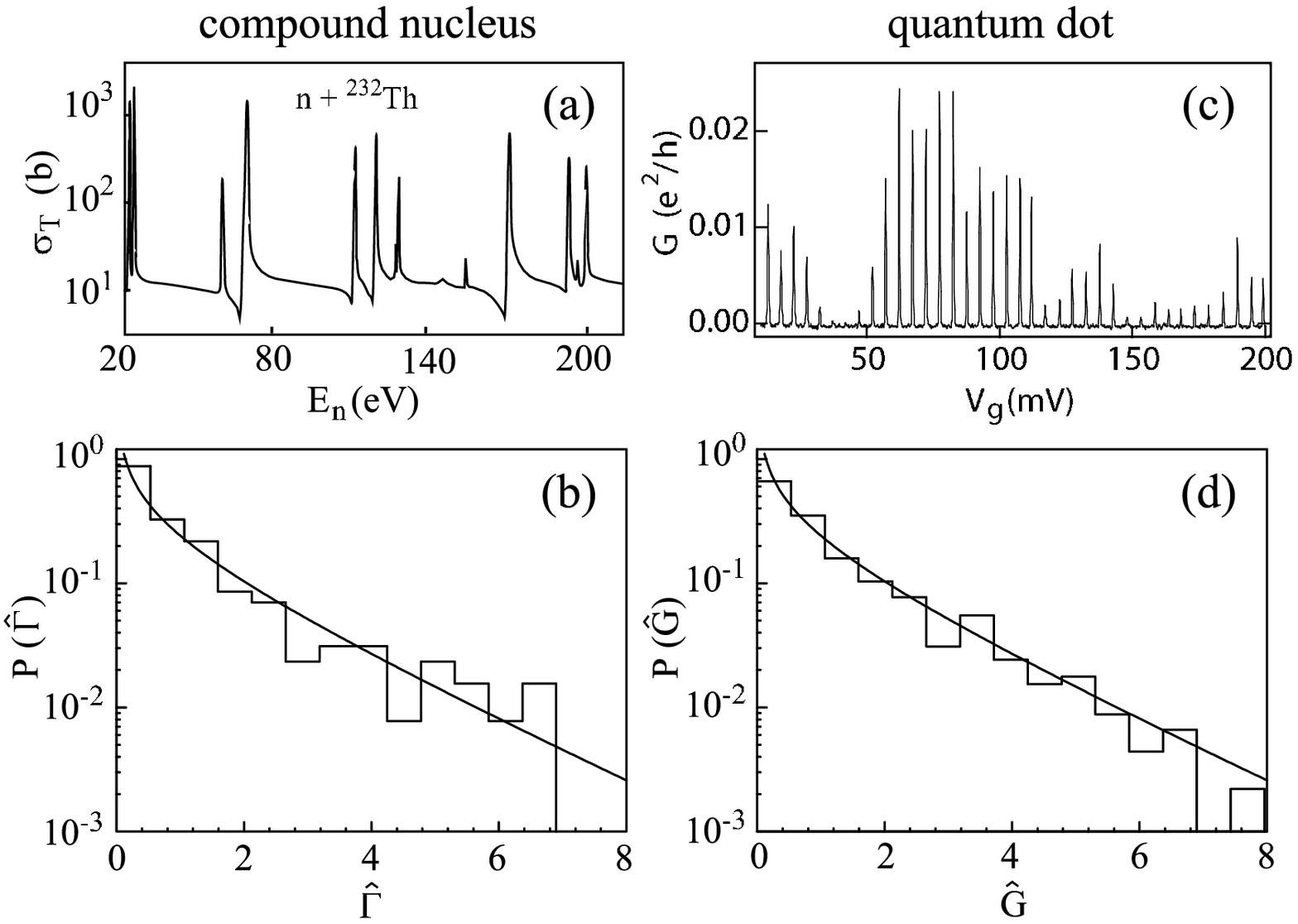}
  \caption{Neutron resonances in a compound nucleus [panels (a) and (b)] and conductance peak heights in a chaotic quantum dot [panels (c) and (d)]: (a) total cross section of $n + ^{232}$Th as a function of the incoming neutron energy $E_n$ (in $eV$); (b) the distribution of the normalized neutron resonance width $\tilde \Gamma =\Gamma/\bar\Gamma$ on a log-linear scale. The histogram is the data of 223 resonances \cite{garg64} and the solid line is the Porter-Thomas distribution $P(\tilde \Gamma) \propto \tilde \Gamma^{-1/2}e^{-\tilde\Gamma/2}$; (c) a series of Coulomb blockade peak in a quantum dot versus gate voltage $V_g$ at zero magnetic field \cite{folk96}; (d) the distribution of the normalized conductance peak heights $\tilde G = G/\bar G$. The histogram is the data of 600 peak heights and the solid line is the RMT prediction of Ref.~\cite{jalabert92}. From Ref.~\cite{alhassid00}.
}
\end{figure}

  For a chaotic dot, the conductance peak height exhibits large fluctuations as a function of the gate voltage (see, e.g., in Fig.~\ref{cn-dot}(c)). In
Ref.~\cite{jalabert92} it was suggested that these fluctuations can be described quantitatively using random matrix theory (RMT). RMT was introduced by Wigner \cite{wigner51} in the 1950's to explain the fluctuation properties of the neutron resonances in the compound nucleus, and further developed by Dyson \cite{dyson62}, Mehta and others. The Hamiltonian matrix is assumed to be random except that it must satisfy the underlying space time symmetries.

 In RMT we have an ensemble of Hamiltonians $H$ described by a Gaussian distribution law
\begin{equation}
 P\left[H\right] \propto
 \exp\left(-{\beta\over{2a^2}} {\rm Tr} H^2 \right) \;,
 \end{equation}
 where $a^2$ is a typical variance of a matrix element of $H$.  There are three types of ensembles: (i) Gaussian orthogonal ensemble (GOE) of real symmetric matrices, (ii) Gaussian unitary ensemble  (GUE) of complex hermitean matrices, and (iii) Gaussian symplectic ensemble (GSE) of real quaternions. These three ensembles are labelled by $\beta=1,2$ and $4$, respectively. The GOE corresponds to conserved time-reversal symmetry while the GUE corresponds to broken time-reversal symmetry (e.g., in the presence of an orbital magnetic field). The GSE is relevant in the presence of spin-orbit scattering and will not be discussed here.

  In nuclei, RMT is used to describe the statistical fluctuations of the many-body levels and eigenstates at finite excitation energies \cite{brody81,zelevinsky96}. This approach was justified by the complexity of the nuclear Hamiltonian. In chaotic quantum dots, RMT is used to describe the statistical fluctuations of the single-particle levels and eigenstates. This is based on the Bohigas-Giannoni-Schmit conjecture \cite{BGS84} stating that the statistical quantal fluctuations of a classically chaotic system are described by RMT.  The universality of RMT holds for time scales $t > \tau$, where
\begin{eqnarray}\label{rmt-univ}
\tau = \left\{ \begin{array}{ll}
  \tau_f  &  \mbox{(chaotic ballistic dot)} \\
\tau_{eq} & \mbox{(compound nucleus)}  \;.
\end{array} \right.
 \end{eqnarray}

  The ballistic Thouless energy of a quantum dot is the energy scale associated with $\tau_f$, i.e., $E_T = \hbar/\tau_f$. The dimensionless Thouless conductance $g_T$ is then defined as the number of single-particle levels within a Thouless energy window
\begin{equation}
g_T= {E_T \over d } \;.
\end{equation}
Estimating $\tau_f \sim L/v_F$ (where $v_F$ is the Fermi velocity) we find $g_T \sim \sqrt{N}$.   According to (\ref{rmt-univ}), the RMT description is valid within $g_T$ levels around the Fermi energy.

\begin{figure}[h!]\label{peaks}
  \includegraphics[height=.33\textheight]{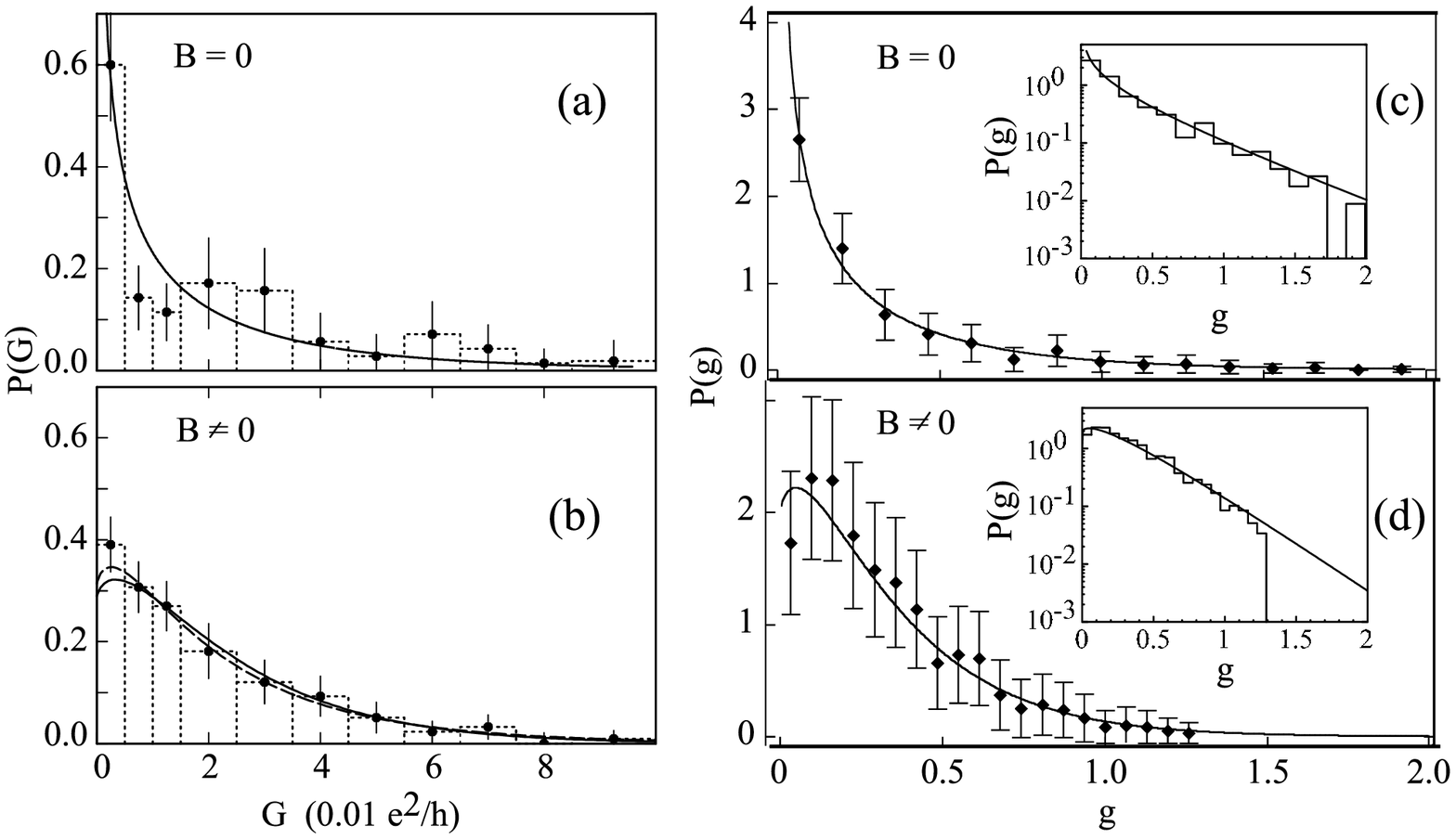}
  \caption{Conductance peak height distributions in Coulomb blockade quantum dots. Left: the measured distributions of Ref.~\cite{chang96} for small quantum dots in the absence [panel (a)] and in the presence [panel (b)] of a time-reversal symmetry-breaking magnetic field $B$.  Right: the measured distributions of Ref.~\cite{folk96} for large quantum dots. The insets show the same distributions (histograms) on a log-linear scale.  The solid lines in all panels are the theoretical predictions of
Ref.~\cite{jalabert92} given by Eqs. (\ref{PgGOE}) and (\ref{PgGUE}) in the absence and presence of a magnetic field, respectively}
\end{figure}

 The distributions of the conductance peak heights in almost-isolated chaotic quantum dots were predicted using RMT \cite{jalabert92}.  The partial width amplitudes correspond to different components of an RMT eigenvector. Thus each of the partial widths $\Gamma^{l,r}$ is described by a Porter-Thomas distribution of the corresponding symmetry $\beta$. Assuming the leads to be uncorrelated, the following closed-form expressions are found for the peak height distributions
\cite{jalabert92,prigodin93}:
\begin{eqnarray}
P_{\rm GOE}(g) & = &  \sqrt{{2 / \pi g}}  e^{-2 g} \label{PgGOE} \\
P_{\rm GUE}(g) & = & 4 g  e^{-2g}  \left[K_0(2g) +K_1(2g) \right]
\;, \label{PgGUE}
\end{eqnarray}
where $K_0$ and $K_1$ are modified  Bessel functions. The distributions  (\ref{PgGOE}) and (\ref{PgGUE}) are universal and depend only on the symmetry class.  They were measured in two independent experiments \cite{folk96,chang96} and found to agree with the theoretical predictions. The comparison between theory and experiment is shown in Fig.~\ref{peaks}.

\vspace{3 mm}

  It is interesting to compare the statistics of the conductance peak heights in an almost-isolated quantum dot with the statistics of the  neutron resonance widths in a compound nucleus. The top panels of Fig.~\ref{cn-dot} compare a typical series of neutron resonances in a compound nucleus (left) with a typical series of conductance peak heights in a quantum dot (right). The bottom panels of Fig.~\ref{cn-dot} compare the distributions of the neutron resonance widths (left) with the distribution of the conductance peak heights in the absence of magnetic field (right).

 Other signatures of quantum chaos in the conductance peak height statistics which we do not discuss here are the parametric peak height correlations (e.g., at different values of the magnetic field) \cite{alhassid96} and the increase of the average conductance with orbital magnetic field \cite{alhassid98}.  They were measured in Refs. \cite{folk96} and \cite{folk01}, respectively.

\subsection{Limitations of the constant interaction model}\label{limitations}

  While the CI plus RMT model was successful in explaining (at least qualitatively) some of the observed mesoscopic fluctuations of the conductance there have also been major discrepancies.
One such discrepancy was observed in the statistics of the peak spacing $\Delta_2$ \cite{sivan96,simmel97,patel98a,luscher01}. At low temperatures, when only the ground state contributes to the linear conductance, the peak spacing is given by
\begin{equation}
\Delta_2 = E_{\rm gs}(N+1) + E_{\rm gs}(N-1) - 2 E_{\rm gs}(N)\;,
\end{equation}
where $E_{\rm gs}(N)$ is the ground-state energy of the dot with $N$ electrons.

\begin{figure}[h!]\label{spacing-ex}
  \includegraphics[height=.23\textheight]{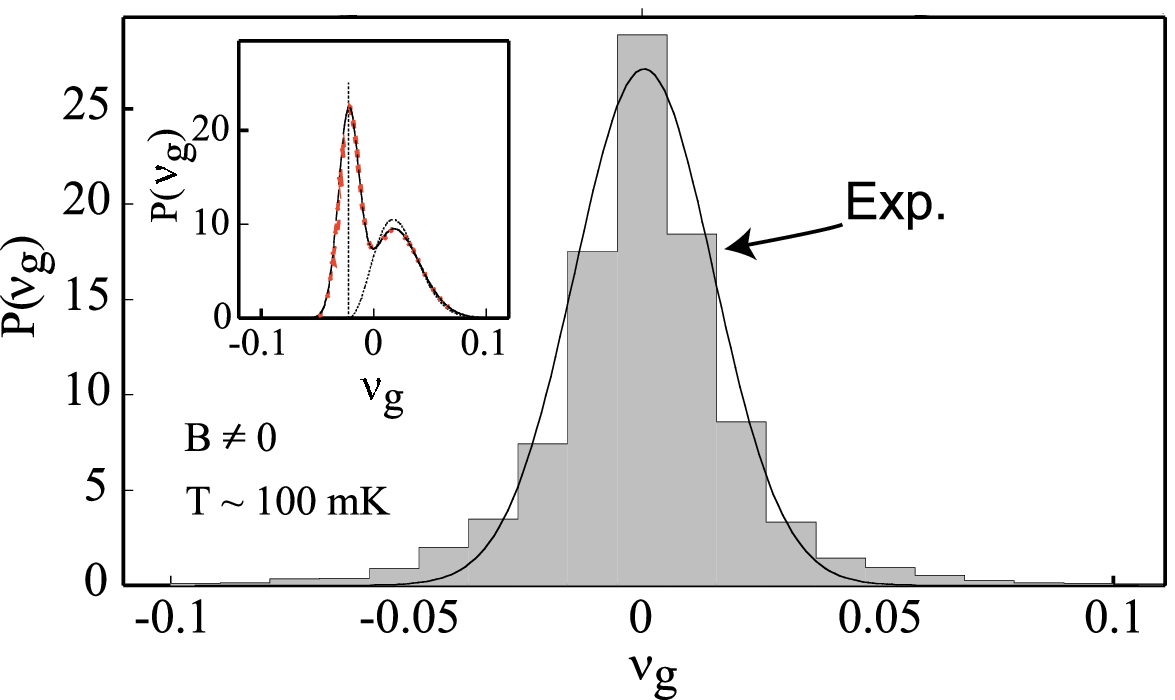}
  \caption{The experimental peak spacing distribution $P(\nu_g)$ where $\nu_g = (\Delta_2 - \bar\Delta_2)/\bar \Delta_2$ \cite {patel98a}. The histogram is the data containing $10800$ peaks at $B \neq 0$ and $k T \sim 0.8 \, d$, and the solid line is a Gaussian fit. The inset is the peak spacing distribution in the CI plus RMT model before (dotted line) and after (solid line) convolution with a Gaussian noise. From Ref.~\cite{patel98a}}
\end{figure}

 In the CI model, $\Delta_2 = e^2/C + \Delta \epsilon$, where $\Delta\epsilon$ is the spacing between two successive single-particle levels [see Eq. (\ref{peak-spacing})]. The charging energy $e^2/C$ is a constant and the statistics of $\Delta_2$ are determined by $\Delta \epsilon$. Since the single-particle levels are spin-degenerate, we expect the peak spacing distribution to be bimodal, i.e., a superposition of a $\delta$-function (corresponding to $\Delta \epsilon=0$) and the Wigner nearest-neighbor spacing distribution (derived from RMT). Including noise, we expect a bimodal distribution shown by a solid line in the inset of Fig.~\ref{spacing-ex}. However, the observed distribution of
Ref.~\cite{patel98a} shown in Fig.~\ref{spacing-ex}) is not bimodal and its shape is closer to a Gaussian. The absence of bimodality is a common observation of all the corresponding experiments \cite{sivan96,simmel97,patel98a,luscher01}.

\vspace{3 mm}

 The peak spacing statistics has also been studied as a function of temperature \cite{patel98a}. The observed width $\sigma(\Delta_2)$ of the spacing fluctuations (symbols in
Fig.~\ref{exchange-s}) is seen to be suppressed when compared with the CI plus RMT model (long dashes).  More detailed studies of the peak height statistics at finite temperature  \cite{patel98b} also reveal significant discrepancies. The $k T=0.1\, d$ experimental distribution, shown by the shaded histogram in the left panel of
Fig.~\ref{exchange-h}, suggests that the CI plus RMT model (dashed histogram) overestimates the probability to find small peak heights. Furthermore, the ratio $\sigma(g_{\rm max})/\bar g_{\rm max}$ between the width and average of the peak height distribution is suppressed in comparison with the predictions of the CI plus RMT model (see right panel of Fig.~\ref{exchange-h}).

\subsection{The universal Hamiltonian}\label{universal-H}

  The origin of the discrepancies discussed in Sec.~\ref{limitations} is in the oversimplified treatment of electron-electron interactions by the CI model. In an almost-isolated dot, charge is quantized and it is necessary to consider correlations beyond the simple charging energy term. Of particular importance are correlations that are associated with the spin degrees of freedom.

\begin{figure}[h!]\label{pauli-hund}
  \includegraphics[height=.23\textheight]{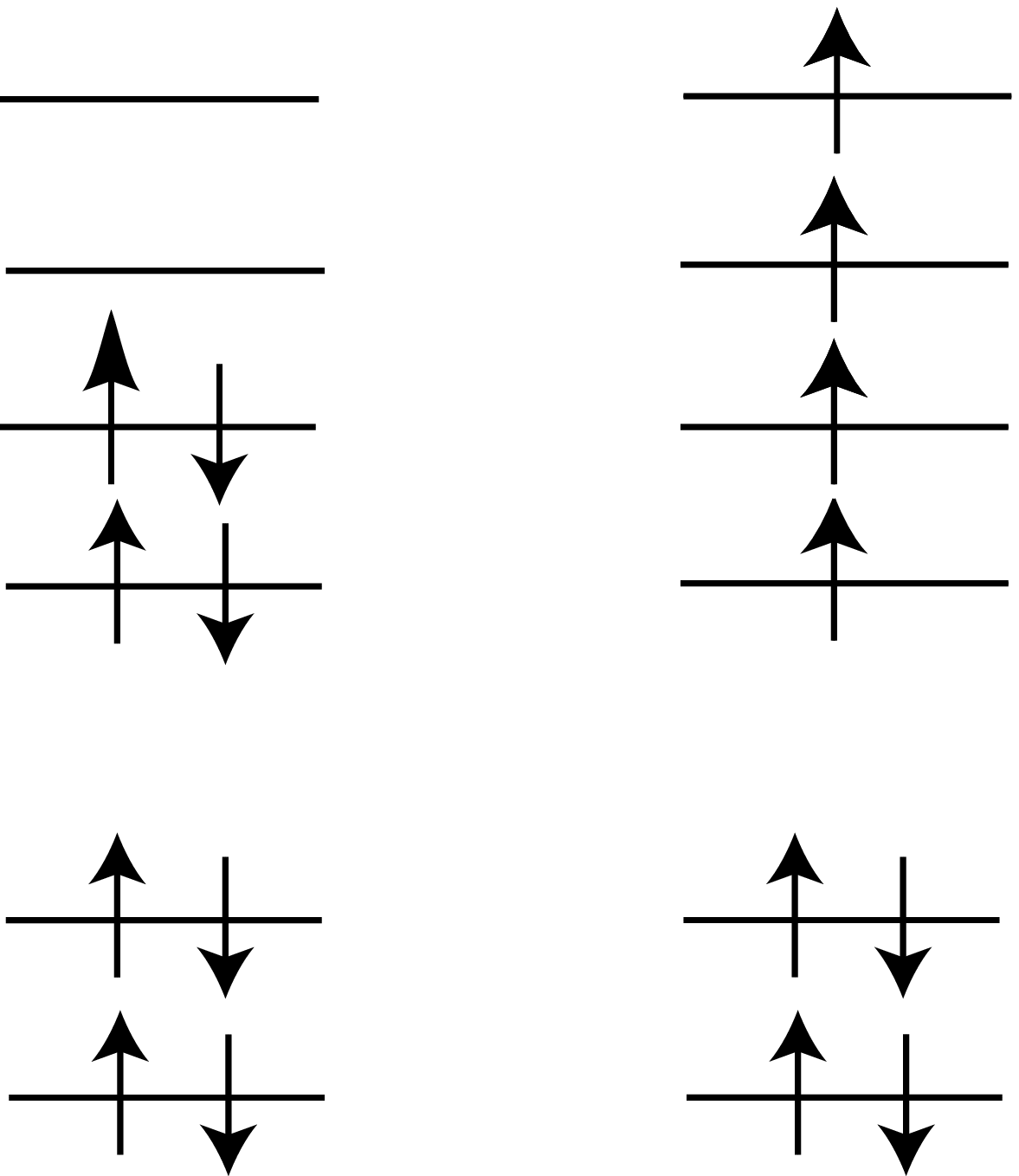}
  \caption{Left: Pauli state, typical of metals. Right: Hund state, typical of atoms.
 }
\end{figure}

   In the absence of interactions, the energy is minimized in a state of lowest spin, i.e., $S=0$ for an even number of electrons and $S=1/2$ for an odd number of electrons. Any configuration with a higher spin requires the promotion of an electron to a higher single-particle level and therefore an increase in energy. Such a state of minimal spin (shown on the left in Fig.~\ref{pauli-hund}) is known as a Pauli state and is typical of metals.

\vspace{3 mm}

 The Coulomb interaction is repulsive and is minimized in a spatially antisymmetric state. Such a state requires a spin-symmetric state to ensure antisymmetrization of the total wave function, i.e., a state of maximal spin. This effect can be described by an effective exchange interaction which is minimized in a state of maximal spin. In atoms, such a state is known as the Hund state (shown on the right in Fig.~\ref{pauli-hund}).  In general, the ground-state spin of a quantum dot is determined by the competition between the total non-interacting energy of the electrons in the confining potential and the exchange interaction.   In chaotic quantum dots in which shell structure is absent, the ground state is often intermediate between a Pauli state and a Hund state.

  In a chaotic dot, the randomness of the single-particle wave functions induces randomness in the interaction matrix elements. In general one can separate the electron-electron interaction into an average and fluctuating parts. The standard deviation of the fluctuating part behaves as $\sim d /g_T$ \cite{blanter97,mirlin00} for a diffusive dot and  $\sim d \sqrt{\ln g_T}/g_T$ for a ballistic dot \cite{blanter01,usaj01,alhassid02}. In the limit of a large Thouless conductance (i.e., for a large number of electrons) only the $g_T$-independent part of the average interaction survives.  This leads to the so-called universal Hamiltonian \cite{kurland00,aleiner02}
\begin{equation}\label{universal}
H_{\rm universal} = \sum_{\lambda \sigma} (\epsilon_\lambda - e\alpha V_g) a^\dagger_{\lambda \sigma} a_{\lambda \sigma} + {e^2 \hat N^2 \over 2 C} - J_s {\bf S}^2 + \delta_{\beta 1} J_c T^\dagger T  \;,
\end{equation}
where $\bf \hat S= {1\over 2} \sum_\alpha
\sum_{\sigma \sigma'} a^\dagger_{\alpha \sigma}
{\hat {\bf \sigma}}_{\sigma \sigma'} a_{\alpha \sigma'}$ is the total spin of the dot (${\hat {\bf \sigma}}$ is the vector of
the three Pauli matrices) and $T^\dagger = \sum_\lambda a^\dagger_{\lambda +} a^\dagger_{\lambda -}$ creates  pairs of spin up-spin down electrons.
$J_s = \bar v_{\alpha\beta; \beta\alpha}$ is the average strength of an exchange interaction matrix element while $J_c = \bar v_{\alpha\alpha; \beta\beta}$ describes an average Cooper channel interaction. The Cooper channel term is allowed only in the orthogonal symmetry (i.e., in the absence of an external magnetic field). In quantum dots, it is repulsive and does not lead to any instability \cite{aleiner02}. In the following discussions on quantum dots we will ignore this term. However, in nanoparticles and nuclei the pairing interaction is attractive and plays an important role, as we shall discuss in Sec.~\ref{nanoparticles} and \ref{nuclei}.  Note that the universal Hamiltonian reduces to the CI model when $J_s=0$ and $J_c=0$.

    The statistical fluctuations of the single-particle wave functions and energies within a Thouless energy band are still described by RMT.  The new important term in the universal Hamiltonian is the constant exchange interaction, and the various statistical properties of the dot will depend on the particular value of the exchange interaction constant $J_s$. RPA estimates provides an expression for $J_s$ as a function of the gas constant $r_s$.  A typical value of $r_s$ for the dots used in the experiments is $r_s \sim 1.2$ and the corresponding RPA estimate is $J_s \sim 0.3 \, d$.

\begin{figure}[h!]\label{spin-prob}
  \includegraphics[height=.25\textheight]{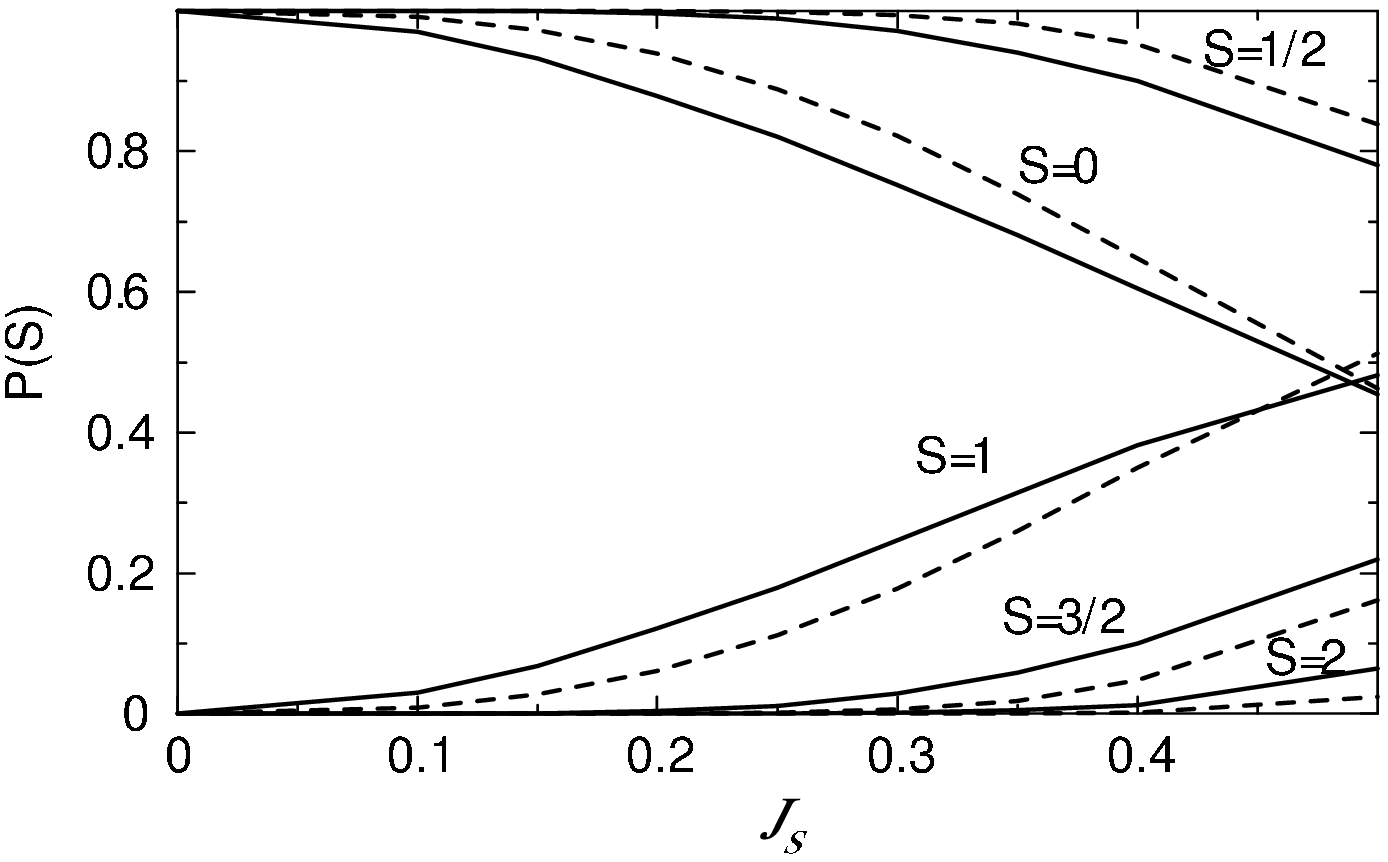}
  \caption{Probability of the ground state to have spin $S$  versus the exchange constant $J_s$ (measured in units of $d$). Solid (dashed) lines correspond to the orthogonal (unitary) symmetry. From Ref.~\cite{alhassid02}.}
\end{figure}

  In the universal Hamiltonian plus RMT model, there is a certain probability for the quantum dot to have spin $S$ in its ground state \cite{oreg02,alhassid02}.  In
Fig.~\ref{spin-prob} we show these ground-state probabilities for different spin values as a function of $J_s$ (measured in units of $d$). In the CI plus RMT model the ground state for an even number of electrons is always characterized by $S=0$. However, for a realistic value of the exchange interaction $J_s=0.3\, d$, the ground-state probability of spin $S=1$ is about 20\%.

\vspace{3 mm}

\subsection{Conductance in the presence of interactions}\label{conductance-int}

    The conductance formula (\ref{g-CI}) is based on a non-interacting model (except for the charging energy term which is constant for a fixed number of electrons).  In the presence of interactions, it is necessary to generalize the rate equation approach using the many-body wave eigenfunctions of the dot. This was accomplished in Ref.~\cite{alhassid04}. The input parameters are the energies of the many-body eigenstates of the dots with $N$ and $N+1$, respectively, and the transition widths $\Gamma^{l(r)}_{ij}$ between the $N$-electron dot in an  eigenstate $i$ and the $N+1$-electron dot in an eigenstate $j$ involving an electron tunneling from the left (right) lead. In the linear response regime it is generally necessary to solve a linear set of equations but in special cases a closed formula can be derived \cite{alhassid04}
\begin{equation}\label{conductance}
G = \frac{e^2}{\hbar kT} \sum_{ij}
{\tilde P}_i^{(N)} f_{ij} \frac{\Gamma_{ij}^{\rm l}
\Gamma_{ij}^{\rm r}}{\Gamma_{ij}^{\rm l} + \Gamma_{ij}^{\rm r}} \;.
\end{equation}
Here $f_{ij}$ is a Fermi-Dirac distribution for an energy-conserving transition between the many-body states $i$ and $j$ and ${\tilde P}_i^{(N)}$ is the equilibrium probability of finding the dot in the eigenstate $i$ of $N$ electrons.

\subsubsection{Conductance formula for a constant exchange interaction}\label{conductance-ex}

 In particular, a closed formula of the form (\ref{conductance}) holds when the orbital occupations $n_\lambda = n_{\lambda +} + n_{\lambda -}$ are good quantum numbers \cite{alhassid04,usaj01}. This is true for the universal Hamiltonian (\ref{universal}) since $[{\bf \hat S}^2, n_\lambda]=0$. However, Eq.~(\ref{conductance}) requires a summation over all the many-body eigenstates of the dot with $N$ and $N+1$ electrons. The number of these eigenstates increases combinatorially with the number of single-particle levels, and (\ref{conductance}) becomes impractical at higher temperatures. Using spin projection techniques we were able to perform the combinatorial sum over all orbital occupation numbers in a closed form \cite{alhassid03}.

  The eigenstates of (\ref{universal}) (with $J_c=0$) are characterized by the orbital occupation numbers ${\bf n} = (n_1,n_2,\ldots)$ of the single-particle levels. The manifold of all Slater determinants with $m$ given singly occupied levels is of dimension $2^m$. In the absence of exchange, all these $2^m$ states are degenerate with energy $\sum_\lambda \epsilon_\lambda n_\lambda$ (not including charging energy).   States with good spin $S$ can then be constructed by taking the appropriate linear combinations of these $2^m$ states. The problem is equivalent to the coupling of $m$ spin-$1/2$ particles to total spin $S$. In general, there are several degenerate states with the same spin $S$ and they can be distinguished by additional quantum numbers $\gamma$.  The eigenstates of good spin $S$ and spin projection $S_z=M$  are thus labeled by $|N {\bf n} \gamma S M\rangle$. Their corresponding energies are $\varepsilon_{{\bf n} S}^{(N)}= \sum_\lambda \epsilon_\lambda n_\lambda+e^2 N^2/2C - J_s S(S+1)$.

  Applying spin (and particle-number) projection, the linear dimensionless conductance can be written as
\begin{equation}\label{g-exchange}
g = \sum_\lambda (w_\lambda^{(0)} + w_\lambda^{(1)}) g_\lambda \;,
\end{equation}
where $g_\lambda$ are the single-level conductances (\ref{g-CI}). The thermal weights  $w_\lambda^{(0)}$ and $w_\lambda^{(1)}$ collect the contributions from processes in which an electron is added to an empty level or to a singly occupied level, respectively. They are given by
\begin{equation}
w_\lambda^{(0)} = 4\sum_S b_{\lambda,N,S} P_{N,S}\!\!\!
\sum_{S'=S\pm 1/2}\!\!\! (2S'\!+\!1) f(\epsilon_{S'S}^\lambda)\,,
\label{w0_spinproj}
\end{equation}
\begin{equation}
w_\lambda^{(1)} = 4\sum_{S'} c_{\lambda,N+1,S'} P_{N+1,S'}\!\!\!
\sum_{S=S'\pm 1/2}\!\!\! (2S\!+\!1) [1\!-\!f(\epsilon_{S'S}^\lambda)]\,,
\label{w1_spinproj}
\end{equation}
where $\varepsilon_{S'S}^\lambda = \varepsilon_{{\bf n'}S'}^{(N+1)} -
\varepsilon_{{\bf n}S}^{(N)} - {\tilde \epsilon}_{\rm F}$. The quantity
$P_{N,S}$ is the probability to find the dot with $N$ electrons and spin $S$. It is given by
 \begin{equation}
P_{N,S} = e^{-\beta [F_{N,S} + U_{N,S}]}/ Z \;,
\end{equation}
where $F_{N,S} = -\beta^{-1} \ln {\rm Tr}_{N,S} e^{-\beta
  \sum_{\lambda\sigma} \epsilon_\lambda \hat n_{\lambda\sigma}}$ is the
 free energy of $N$ non-interacting electrons with total spin $S$,
 $U_{N,S} = e^2 N^2/2C - J_s S(S+1) - \tilde \epsilon_{\rm F} N$ and $Z$ is the grand-canonical partition function restricted to $N$ and $N+1$ electrons.
The quantities $b_{\lambda,N,S}= \frac{1}{2} \langle({\hat n}_\lambda\!-\!1) ({\hat n}_\lambda\!-\!2)\rangle_{N,S}$ and $c_{\lambda,N,S}= \frac{1}{2} \langle {\hat n}_\lambda ({\hat n}_\lambda\!-\!1) \rangle_{N,S}$ are projected quantities at fixed particle-number $N$ and spin $S$.

  The free energy $F_{N,S}$ and the coefficients $b,c$ can be expressed in closed form in terms of the free energy ${\tilde F}_q$ and canonical occupations $\langle\tilde n_\lambda\rangle_q$ of $q$ spinless fermions with
energies $\epsilon_\lambda$:
\begin{equation}\label{sp_freeenergy}
e^{\!\!-\beta F_{N,S}}\!=\!e^{\!\!-\beta({\tilde F}_{\frac{N}{2}+S}\!+{\tilde
F}_{\frac{N}{2}-S})} \!\!\!-\! e^{\!\!-\beta({\tilde F}_{\frac{N}{2}+S+1}\!
+{\tilde F}_{\frac{N}{2}-(S+1)})}\!\!.
\end{equation}
\begin{equation}\label{c-function}
c_{\lambda,N,S} = \frac{\langle {\tilde n}_\lambda
\rangle_{\frac{N}{2}+S} \langle {\tilde n}_\lambda \rangle_{\frac{N}{2}-S}
e^{-\beta (\tilde F_{\frac{N}{2}+S} + \tilde F_{\frac{N}{2}-S})}
- \langle {\tilde n}_\lambda \rangle_{\frac{N}{2}+S+1}
\langle {\tilde n}_\lambda \rangle_{\frac{N}{2}-(S+1)}
e^{-\beta (\tilde F_{\frac{N}{2}+S+1} + \tilde F_{\frac{N}{2}-(S+1)})}}
{e^{-\beta(\tilde F_{\frac{N}{2}+S} +\tilde  F_{\frac{N}{2}-S})} -
e^{-\beta(\tilde F_{\frac{N}{2}+S+1} + \tilde F_{\frac{N}{2}-(S+1)})}} \;.
\end{equation}
$b_{\lambda,N,S}$ is found is by replacing $\tilde n_\lambda$ in Eq.~(\ref{c-function}) with $(1-\tilde n_\lambda)$.

 For a finite number of single-particle levels, the quantities ${\tilde F}_q$ and $\langle\tilde n_\lambda\rangle_q$ can be expressed in closed form using particle-number projection [see, e.g., Eqs. (140) in Ref.~\cite{alhassid00}].

\subsubsection{Peak spacing statistics}

\begin{figure}[h!]\label{exchange-s}
  \includegraphics[height=.27\textheight]{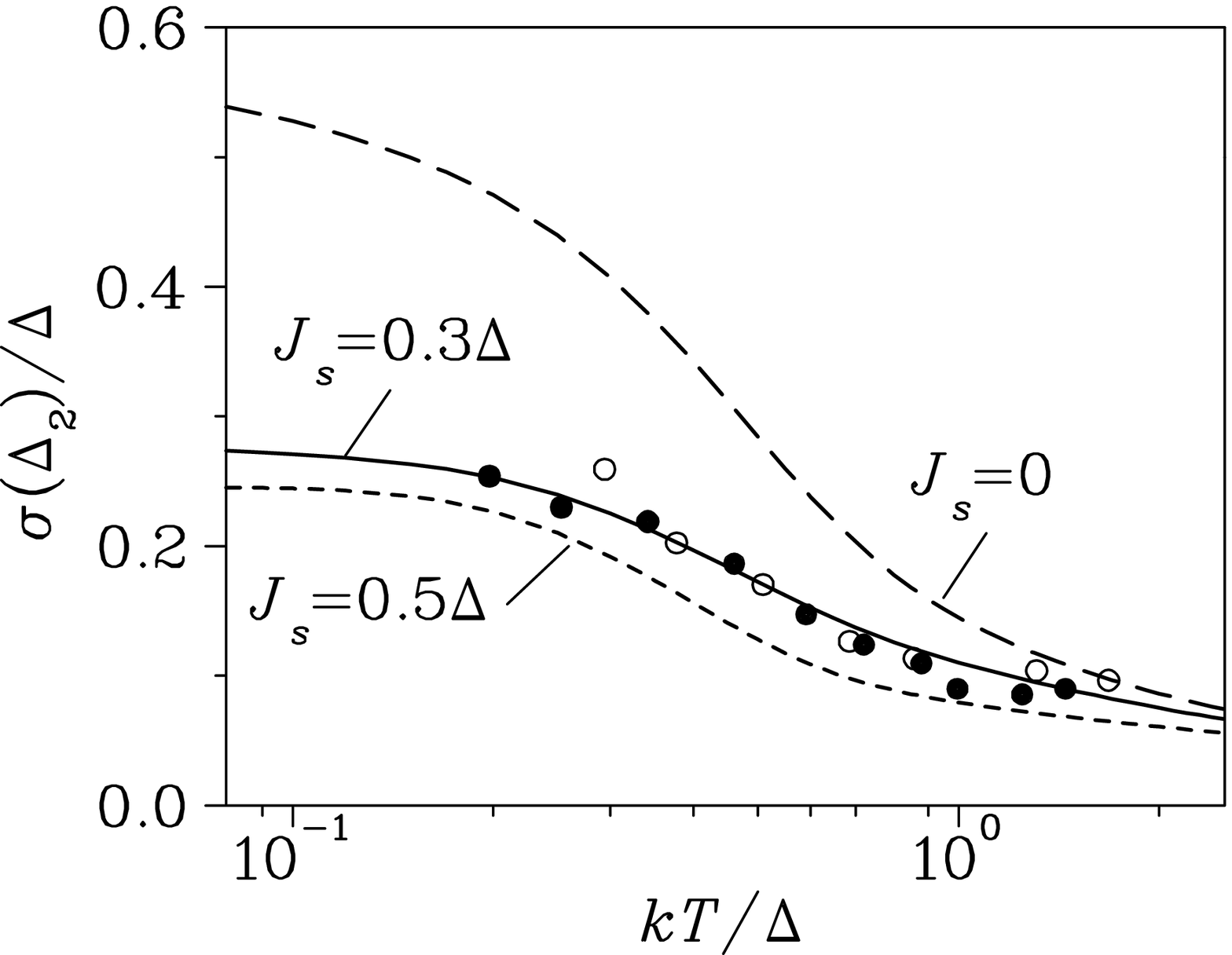}
  \caption{Standard deviation of peak spacing fluctuations $\sigma(\Delta_2)$ in the presence of magnetic field as a function of $k T/d$. Lines are theoretical results for $J_s=0, 0.3\, d$ and $0.5\, d$ and the symbols are the experimental results of Ref.~\cite{patel98a}. From Ref.~\cite{alhassid03}.}
\end{figure}

In Ref.~\cite{alhassid03}, we have used the conductance formula of the universal Hamiltonian (discussed in Sec.~\ref{conductance-ex}) to calculate the statistics of peak spacings and peak heights. For a chaotic dot, the single-particle spectrum and the partial widths $\Gamma_\lambda$ are calculated from random matrices with the appropriate symmetry. For each sample, the conductance is calculated as a function of $\tilde \epsilon_F$ using the formulas of Sec.~\ref{conductance-ex} and then maximized to find the peak height.   Figure \ref{exchange-s} shows the width $\sigma(\Delta_2)$ of the peak spacing fluctuations as a function of $kT/d$. The solid line corresponds to $J_s=0.3\, d$ and is in excellent agreement with the experimental results (symbols). We note that this value of $J_s$ is not fitted but is extracted from the value of the gas constant $r_s$ of the experimental samples using an RPA estimate.  When compared with the CI model ($J_s=0$, long dashed), we observe that the exchange interaction suppresses the fluctuations of the peak spacing.

\subsubsection{Peak height statistics}

\begin{figure}[h!]\label{exchange-h}
  \includegraphics[height=.3\textheight]{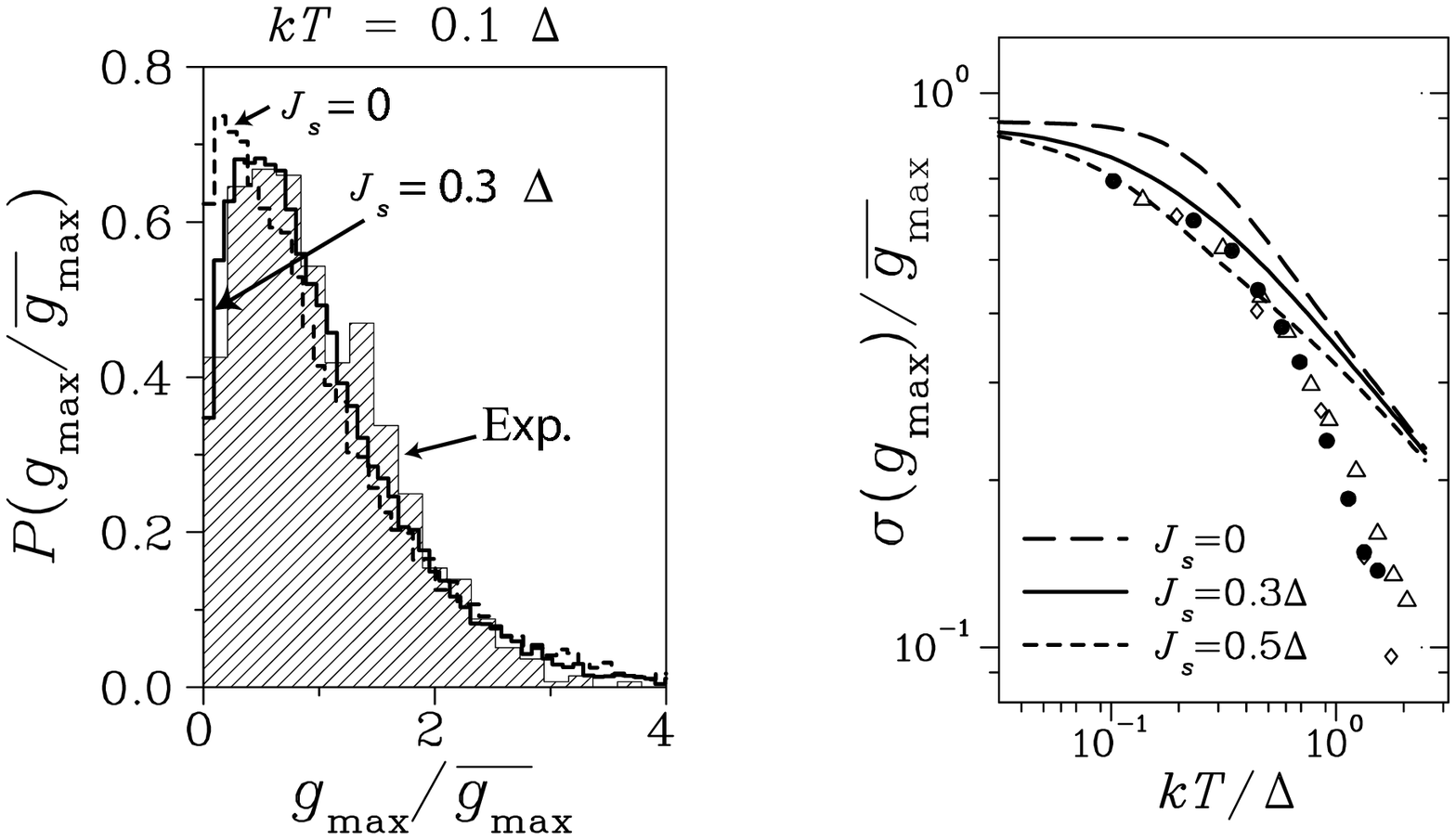}
  \caption{Peak height statistics in chaotic dots (in the presence of magnetic field). Left: the experimental peak height distribution \cite{patel98b} for $k T= 0.1\, d$ (shaded histogram) is compared with the theoretical distributions for $J_s=0$ (dashed histogram) and for $J_s=0.3\, d$ (solid histogram). Right: the ratio $\sigma(g_{\rm max})/\bar g_{\rm max}$ between the standard deviation and the average value of the peak
height as a function of $kT/d$.  Theoretical results are shown for $J_s=0$ (long-dashed line), $J_s=0.3 d$ (solid line) and $J_s=0.5\, d$ (short-dashed line).  The experimental results of Ref.~\cite{patel98b} are shown by symbols. From Ref.~\cite{alhassid03}.}
\end{figure}

 The peak height statistics \cite{alhassid03,usaj03} are shown in Fig.~\ref{exchange-h}. The left panel shows the peak height distribution at the lowest experimental temperature of $k T= 0.1\, d$.  The calculated distribution for $J_s=0.3\, d$
(solid histogram) describes well the experimental distribution (shaded histogram).  On the other hand, the CI model (dashed histogram) overestimates the probability of small peak heights. The ratio $\sigma(g_{\rm max})/\bar g_{\rm max}$ between the average and width of the peak height distribution is shown versus $kT/d$ in the right panel of Fig.~\ref{exchange-h}. This ratio is found to be suppressed by the exchange interaction. The $J_s=0.3\, d$ results (solid line) are in good agreement with the experiment (symbols) for low and intermediate temperatures $kT \leq 0.6\, d$.  At higher temperatures there are deviations that are still not understood.

\vspace{3 mm}

\subsection{Beyond the universal Hamiltonian: residual interaction effects}\label{residual}

  For a finite Thouless conductance $g_T$, it is necessary to take into account the residual interactions beyond the universal Hamiltonian. In general, the effective Hamiltonian of a chaotic dot in a band of $\sim g_T$ levels around the Fermi energy is
\begin{equation}\label{residual-int}
H = \sum_{\alpha \sigma} \epsilon_{\alpha} a^\dagger_{\alpha \sigma}
a_{\alpha \sigma}+ {1 \over 2} \sum_{\alpha \beta \gamma \delta \atop
\sigma \sigma'} v_{\alpha \beta; \gamma \delta} a^\dagger_{\alpha
\sigma} a^\dagger_{\beta \sigma'} a_{\gamma \sigma'} a_{\delta \sigma}\;,
\end{equation}
where $v$ is the two-body screened Coulomb interaction in 2D.

The randomness of the single-particle wave functions induces a randomness of the interaction matrix elements when the latter are calculated in the single-particle eigenstates. The possible induced ensembles of random interactions, which depend on the underlying space-time symmetries and on features of the two-body interactions, were classified in Ref.~\cite{alhassid04a}. The cumulants of the interaction matrix elements are determined by requiring
invariance under a change of the single-particle basis. The average interaction is characterized by three (two) invariants in the orthogonal (unitary) symmetry
$\bar v_{\alpha \beta; \gamma \delta} = v_0 \delta_{\alpha \gamma}
\delta_{\beta \delta} + J_s \delta_{\alpha \delta} \delta_{\beta \gamma}
+J_c \delta_{\alpha \delta} \delta_{\beta \gamma}$, and has the same form of the universal Hamiltonian (except for a linear term in $\hat N$). The second cumulant for a local two-body interaction in the unitary symmetry is given by
\begin{equation}
\overline{ \delta v^*_{\alpha \beta; \gamma \delta} \delta v_{\mu \nu; \rho \sigma}}
= u^2 (\delta_{\alpha \mu} \delta_{\beta \nu} \delta_{\gamma \rho}
\delta_{\delta \sigma} +
\delta_{\alpha \nu} \delta_{\beta \mu} \delta_{\gamma \sigma}
\delta_{\delta \rho}) \;.
\end{equation}
The corresponding expression in the orthogonal symmetry can be found in Ref.~\cite{alhassid04a}. The constant $u^2$ describes a typical variance of a screened Coulomb interaction matrix element, and can be estimated using a contact interaction model in the limit of a small screening radius. In a diffusive dot $u \sim d/g_T$ \cite{blanter97} and in a ballistic dot $\sim d \sqrt{\ln g_T}/g_T$ \cite{blanter01,usaj01,alhassid02}.

  In the experiments, $g_T =\pi \sqrt{N/2}$ varies in the range $\sim 30 - 70$ and the fluctuations of the interaction matrix elements are expected to be small compared with $d$. The effect of the residual interaction terms can be studied within the Hartree-Fock (HF) approximation. Most of these studies has been restricted to spinless electrons \cite{HF,alhassid04a}.

 In a finite dot, screening also leads to a surface charge potential \cite{blanter97} [not included in (\ref{residual-int})]. The fluctuations of the matrix elements of the one-body surface charge  potential are of order $d/\sqrt{g_T}$, parametrically larger than the fluctuations of the two-body interaction matrix elements. However, for the experimental values of $g_T$, the fluctuations of the one-body and two-body interaction matrix elements turned out to be comparable.

\vspace{3 mm}
\begin{figure}[h!]\label{spacing-dist}
  \includegraphics[height=.23\textheight]{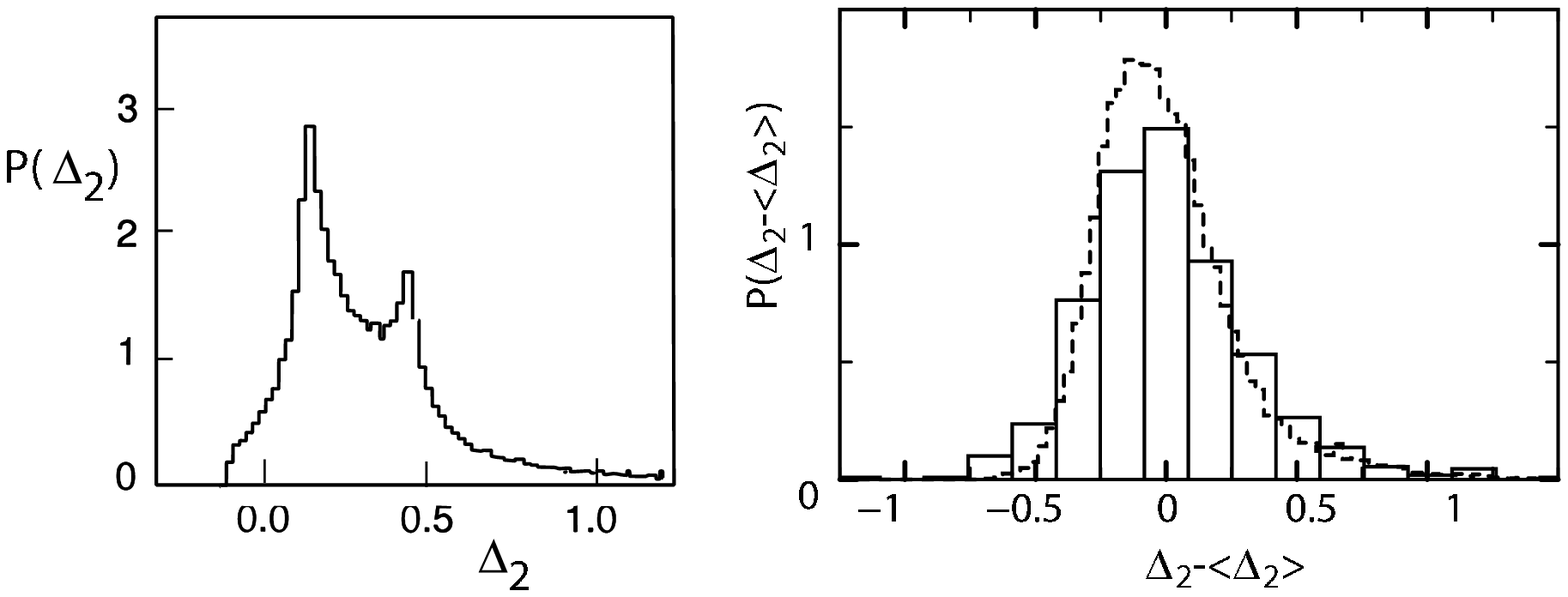}
  \caption{Left: the peak spacing distribution at $kT=0.2\, d$ for the universal Hamiltonian (\ref{universal}) with $J_s=0.3\, d$ (unitary symmetry). Right: the peak spacing distribution at $kT=0.22 d$. The dashed  histogram is the calculated distribution for $J_s=0.28\, d$ and includes residual interaction effects in the HF-Koopmans approach. The solid histogram is the experimental distribution of Ref.~\cite{patel98a}. From Ref. \cite{alhassid02}.}
\end{figure}

  A prominent experimental feature of the peak spacing distribution is the absence of bimodality \cite{sivan96,simmel97,patel98a,luscher01}. For temperatures below $\sim 0.3\, d$, the peak spacing distribution remains bimodal \cite{usaj01} even when the exchange interaction is included (see left panel of Fig.~\ref{spacing-dist}). The  absence of bimodality was understood to be an effect of the residual interactions \cite{berkovits98,ullmo01,usaj01,alhassid02,hirose02}. Residual interactions (in the presence of spin) were studied using different methods including diagonalization for a small number of electrons \cite{berkovits98}, a Strutinsky approach \cite{ullmo01}, spin density functional theory \cite{hirose02}, and a HF-Koopmans approach \cite{alhassid02} in which the single-particle wave functions are assumed to be unchanged upon the addition of electrons \cite{koopmans34}.  In the HF-Koopmans approach, we choose a reference HF many-particle state (e.g., an $S=0$ Pauli state for $N$ electrons) and then express the addition energy in terms of the HF single-particle levels of the reference state and a few interaction matrix elements.  Using the fluctuation properties of the HF levels and wave functions we can determine the residual interaction effects on the peak spacing distribution.
The results (including the effects of surface charge effects and gate voltage scrambling) are shown in the right panel of Fig.~\ref{spacing-dist}. The theoretical peak spacing distribution compares reasonably well with the experimental distribution.

\section{Metallic nanoparticles}\label{nanoparticles}

\begin{figure}[h!]\label{nanoparticle}
  \includegraphics[height=.18\textheight]{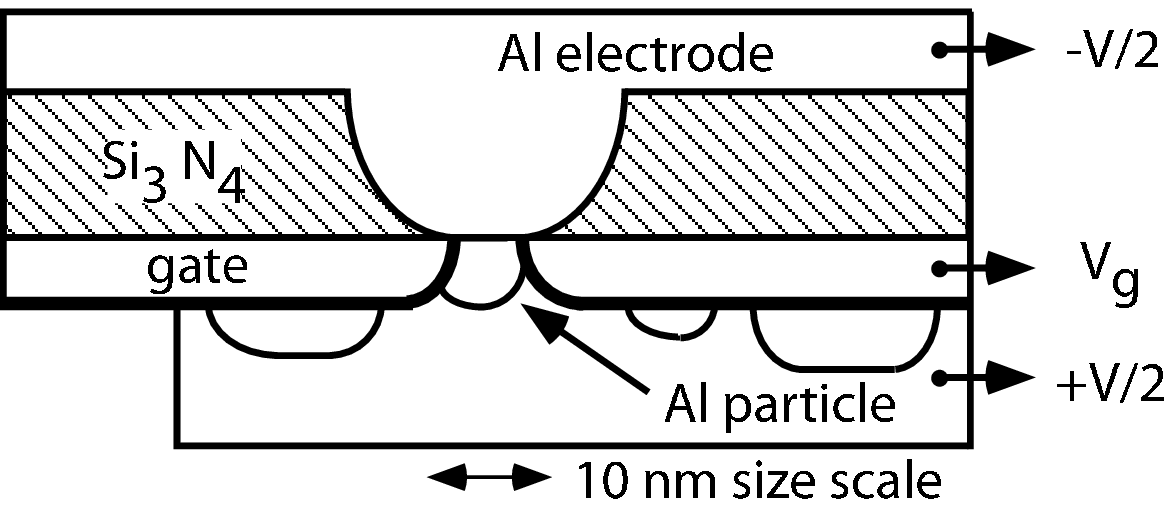}
  \caption{A schematic drawing of the experimental setup of Ref.~\cite{RBT97} for the transport measurements in $Al$ nanoparticles. From Ref.~\cite{RBT97}.}
\end{figure}

   The fabrication techniques for quantum dots were extended in the mid 1990's to ultrasmall metallic grains. Figure \ref{nanoparticle} shows an ultrasmall metallic grain (of size $\sim 2 - 10$ nm) connected to leads \cite{RBT97}.  For such grains the temperature is much smaller than the mean level spacing $d$ and the discreteness of the levels becomes important. By measuring the non-linear conductance it is possible to determine the excitation spectrum of an individual grain.

\vspace{3 mm}

 In contrast to quantum dots, the metallic nanoparticles become superconducting at low temperatures. Experimental and theoretical studies of ultrasmall metallic grains have led to a better understanding of the role of pairing interaction in finite-size systems. For a recent review see Ref.~\cite{nano01}.

  A simple model of a metallic gain is given by
\begin{equation}\label{nanograin}
H_{\rm nanoparticle} = \sum_{\lambda \sigma} \left(\epsilon_\lambda - e\alpha V_g +
{1 \over 2} g_B \mu_B H \sigma \right) a^\dagger_{\lambda \sigma} a_{\lambda \sigma} + {e^2 \hat N^2 \over 2 C} + J_c T^\dagger T  \;.
\end{equation}
with $J_c <0$.  This is essentially the universal Hamiltonian (\ref{universal}) but with an attractive pairing interaction. We have also included a Zeeman energy term in an external magnetic field $H$ ($\mu_B$ is the Bohr magneton and $g_B$ is the $g$-factor).

 The pairing interaction in (\ref{nanograin}) has the reduced BCS form.  BCS theory explained the phenomenon of superconductivity in bulk metals. As a mean field theory it is valid in the limit when the mean level spacing $d$ is much smaller than the bulk pairing gap $\Delta$. In this limit fluctuations can be ignored and a mean field theory is a good approximation.

\begin{figure}[h!]\label{nano-levels}
  \includegraphics[height=.23\textheight]{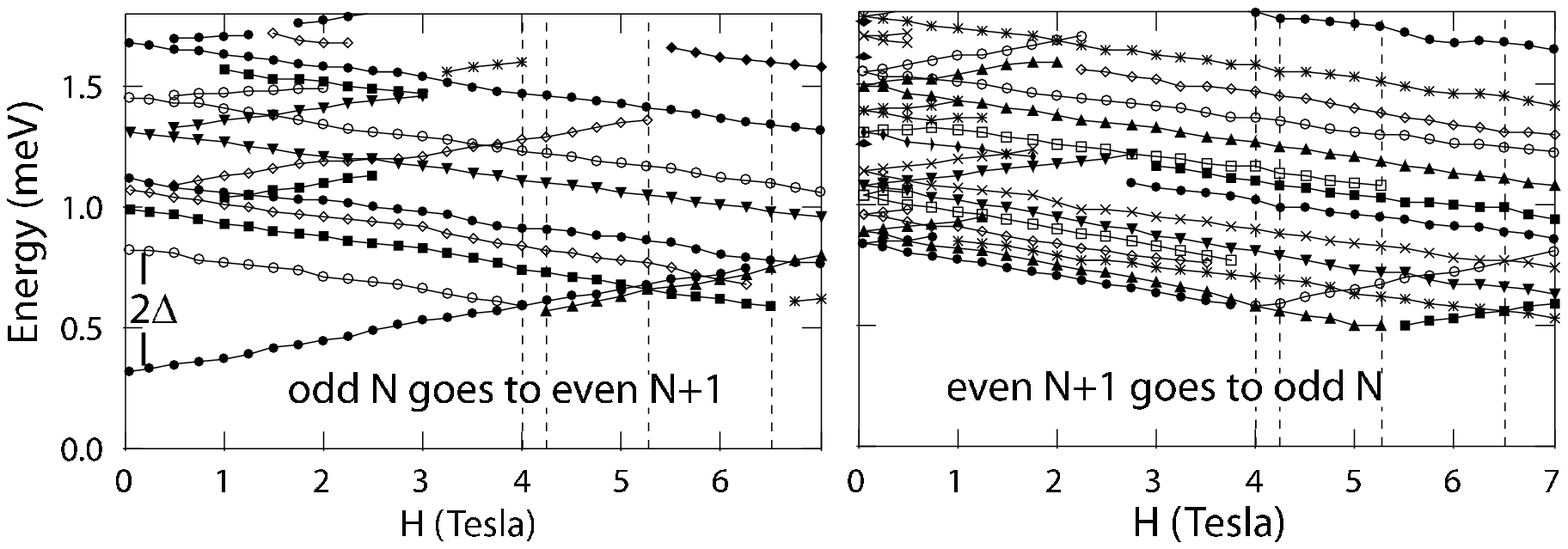}
  \caption{The excitation spectrum of an $Al$ nanoparticle versus a magnetic field $H$ extracted from non-linear conductance measurements. Left: transition from odd to even number of electrons. Notice the gap $2 \Delta$ to the first excited state. Right: transition from even to odd number of electrons. From Ref.~\cite{RBT97}.}
\end{figure}

  This BCS limit holds for metallic nanoparticles that are larger than $\sim 5$ nm. Indeed a pairing gap was observed in the spectrum of such nanoparticles with an even number of electrons. An example is shown in Fig.~\ref{nano-levels} where the excitation spectrum of an $Al$ grain is shown as function of a magnetic field $H$. A gap is observed for an even number (left panel) but not for an odd number of electrons (right panel).  In BCS theory, the low excitations for an even number of electrons involves the excitation of two quasi-particles, hence a gap of $\sim 2\, \Delta$ to the first excited state. On the other hand, the lowest states in a grain with an odd number of electrons are described by one quasi-particle states and hence no gap is observed between the first excited state and the ground state.

  However, in the smaller grains of size less than $\sim 3$ nm, we have $d \geq \Delta$ and the effect seen in
Fig.~\ref{nano-levels} can no longer be observed. This is the fluctuation-dominated regime where BCS theory is no longer valid. An interesting question is whether signatures of pairing correlations still survive in the crossover from BCS to the fluctuation-dominated regime. Theoretical studies have indicated that the crossover is smooth but depends on the particle-number parity (see in \cite{nano01} and references therein). To derive such effects it is important to project on a fixed number of electrons or at the very least project on the correct parity of the number of electrons.  In Sec.~\ref{nuclei} we address pairing correlations in the crossover regime by studying properties of nuclei at finite temperature.

\section{Nuclei}\label{nuclei}

  In a nucleus, a typical spacing $d$ within a shell is smaller than the pairing gap $\Delta$. Thus the nucleus belongs to a regime dominated by fluctuations just as is the case for the smaller nanoparticles.  While zero-temperature signatures of the pairing interaction are well known in nuclei, much less is known about thermal signatures of pairing in nuclei.

 To find out whether thermal signatures of the pairing transition survive in finite nuclei (despite the large fluctuations), it is necessary to take into account correlations beyond the mean field approximation. Interaction effects can be described by including fluctuations around the mean field. This is formally described by the Hubbard-Stratonovich (HS) transformation \cite{HS57}, in which the Gibbs ensemble $e^{-\beta H}$ at inverse temperature $\beta$ is described as a superposition of one-body propagators $U_\sigma$
\begin{equation}\label{HS}
e^{-\beta H} = \int {\cal D}[\sigma]
G_\sigma U_\sigma \;.
\end{equation}
Here $G_\sigma$  is a Gaussian weight and $U_\sigma$ is the imaginary-time propagator describing non-interacting nucleons moving in time-dependent external auxiliary fields $\sigma(\tau)$.

  \subsection{Quantum Monte Carlo methods}\label{SMMC}

   Nuclear properties are well described by the interacting shell model, in which the nucleons move in a mean field potential and interact via residual interactions. The model is usually solved by conventional diagonalization of the Hamiltonian in the many-particle space. The limitation of this method is the combinatorial increase of the model space with the number of nucleons and/or the number of single-particle levels. Shell Monte Carlo (SMMC) methods that are based on the HS decomposition were developed to overcome this limitation \cite{LJK93,ADK94,SMMC}. The  integrand in (\ref{HS}) can be easily calculated using matrix algebra in the single-particle space. However, the number of auxiliary fields $\sigma(\tau)$ is very large and the integration in (\ref{HS}) is done by Monte Carlo methods.

The thermal expectation value of an observable $O$ is given by
\begin{eqnarray}\label{observ}
\langle O \rangle=
{{\rm Tr}\,( O e^{-\beta H})\over{\rm Tr}\,(e^{-\beta H})}=
{\int {\cal D}[\sigma] G_\sigma \langle O \rangle_\sigma{\rm Tr}\,U_\sigma
\over \int {\cal D}[\sigma] G_\sigma {\rm Tr}\,U_\sigma}  \;,
\end{eqnarray}
 where  $\langle O \rangle_\sigma\equiv
 {\rm Tr} \,( O U_\sigma)/ {\rm Tr}\,U_\sigma$ is the expectation value of the observable $O$ evaluated for a sample $\sigma$.  In SMMC we choose the samples according to the probability distribution $W_\sigma\equiv G_\sigma \vert {\rm Tr} U_\sigma \vert$, and the thermal expectation value of the observable $O$
is then calculated estimated from
\begin{equation}
\langle O \rangle \approx { \sum_\sigma \langle O \rangle_\sigma \Phi_\sigma \over \sum_\sigma \Phi_\sigma}
\end{equation}
where $\Phi_\sigma\equiv {\rm Tr} \; U_\sigma /\vert {\rm Tr}\; U_\sigma \vert$
is the sign of the one-body partition function ${\rm Tr}\, U_\sigma$.

 Since the number of nucleons is relatively small, it is important to use the canonical ensemble at fixed number of protons and neutrons. This can be achieved through particle-number projection \cite{ormand94}. For example, the partition function for $A$ particles in $N_s$ single-particle orbitals is given by
\begin{eqnarray}\label{canonical}
{\rm Tr}_A U_\sigma =\frac{e^{-\beta\mu A}
}{N_s}\sum_{m=1}^{N_s}
e^{-i\phi_m  A} \det \left[ {\bf 1}+e^{i\phi_m}e^{\beta\mu}{\bf
U}_\sigma\right]
\;,
\end{eqnarray}
where $\phi_m=2\pi m/N_s \;\; (m=1,\ldots,N_s)$ are quadrature points,  and
$\mu$ is a chemical potential.

  Often the sign $\Phi_\sigma$ fluctuates from sample to sample. When the statistical error of the sign is larger than its average value, the statistical errors of observables become too large and the method fails. This is known as the sign problem and is generic to all fermionic Monte Carlo methods.
  A typical effective nuclear interaction suffers from such a sign problem. In general the interaction can be decomposed into ``good" and ``bad" components, depending on whether a sign problem is absent or not (for the corresponding component).  The dominant collective components of the nuclear interaction are good-sign and this property was used in Ref.~\cite{ADK94} to find a practical solution to the sign problem in finite nuclei.

 \subsection{Pairing correlations at finite temperature}\label{pairing}

 \subsubsection{Hamiltonian}

   We have constructed nuclear interactions that contain only the dominant collective components, i.e., pairing and multipole interactions.  In particular, we have considered the following isospin conserving interaction \cite{NA97}
\begin{eqnarray}\label{nuclear}
  H = \sum_a \epsilon_a \hat n_a - g_0 P^{(0,1)\dagger}\cdot \tilde P^{(0,1)}
     - \chi \sum_\lambda k_\lambda O^{(\lambda,0)}\cdot O^{(\lambda,0)} \;,
\end{eqnarray}
where
\begin{eqnarray}\label{operators}
 P^{(\lambda,T)\dagger}&=&{\sqrt{4\pi}\over{2(2\lambda +1)}}
 \sum_{ab} \langle j_a\| Y_\lambda \|j_b\rangle
 [a_{j_a}^\dagger \times a_{j_b}^\dagger]^{(\lambda,T)}\;, \nonumber\\
O^{(\lambda,T)}&=&{1\over\sqrt{2\lambda +1}}
 \sum_{ab} \langle j_a\| {{dV}\over{dr}} Y_\lambda \|j_b\rangle
[a_{j_a}^\dagger \times \tilde a_{j_b}]^{(\lambda,T)} \;.
\end{eqnarray}
and $\tilde a_{j,m,m_t} = (-)^{j-m+{1\over 2}-m_t} a_{j,-m,-m_t}$ (
(a similar definition is used for $\tilde P^{(\lambda,T)}$). The mean field single-particle orbitals are denoted by $a\equiv(n l j m)$ with principal quantum number $n$, orbital angular momentum $l$ and total angular momentum and projection $j, m$. The corresponding single-particle energies $\epsilon_a$ are calculated in a Woods-Saxon potential plus spin-orbit interaction \cite{BM69}. The one-body  potential $V$ in (\ref{operators}) is the central part of the Woods-Saxon potential, and the multipole interaction is obtained by expanding the surface-peaked interaction $v({\bf r}, {\bf r}^\prime)
 = -\chi (dV/dr)(dV/dr^\prime) \delta(\hat{\bf r} - \hat{\bf r}^\prime)$. In the expansion we only keep the multipoles $\lambda=2,3,4$. The parameter $\chi$ is determined self-consistently \cite{ABDK96} $\chi^{-1} = \int_0^\infty dr \; r^2  \left(dV/ dr \right)  \left(d\rho/ dr
\right)$ ($\rho$ is the nuclear density), and $k_\lambda$ are renormalization factors that take into account polarization of the core (we have used $k_2=2$, $k_3=1.5$ and $k_4=1$). The pairing coupling strength $g_0$ is determined using experimental odd-even mass differences. All interaction components in (\ref{nuclear}) are attractive and lead to a good-sign Hamiltonian. We have studied nuclei in the mass range $A \sim 50 - 70$ using the Hamiltonian (\ref{nuclear}) in the complete $fpg_{9/2}$ shell (containing $30$ orbitals for protons and $30$ orbitals for neutrons).

  The pairing interaction is an important component of the effective nuclear interaction (\ref{nuclear}). In the bulk, BCS theory predicts a finite discontinuity in the heat capacity across the pairing transition temperature. However the nucleus is a finite system and its heat capacity is expected to behave smoothly as a function of temperature. As discussed above, the nucleus is in a regime dominated by fluctuations and an interesting question is whether signatures of the pairing transition survive despite the large fluctuations.

 In the absence of multipole interactions, the Hamiltonian (\ref{nuclear}) has the reduced BCS form, and its eigenvalues can be found by solving a closed set of non-linear equations \cite{richardson}.  However, since we are interested in finite temperature properties, the SMMC approach offers a more direct approach. Furthermore, the SMMC method allows us to treat the more realistic Hamiltonian (\ref{nuclear}).

\subsubsection{Heat capacity}

  In SMMC, the thermal energy $E$ is calculated as a function of inverse temperature $\beta$ and the heat capacity is calculated from a numerical derivative $C=dE/d T$. However, this leads to large statistical errors in the vicinity of the transition temperature (even for a good-sign interaction).  We have introduced a novel method to calculate the heat capacity that reduces the statistical error by about an order of magnitude \cite{LA01}. This method is based on correlated errors.

\begin{figure}[h!]\label{bcs}
  \includegraphics[height=.4\textheight]{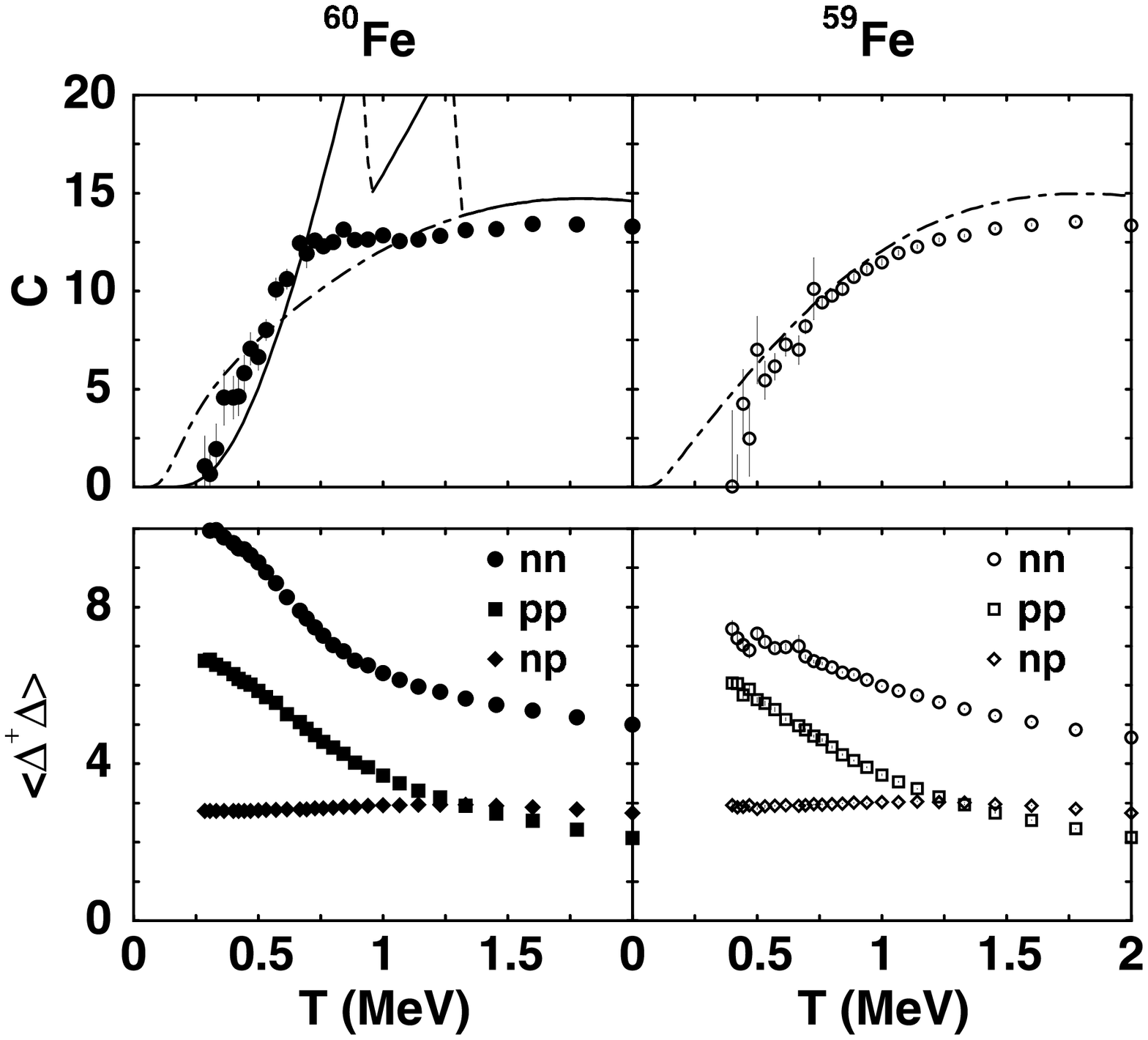}
  \caption{Top panels: heat capacities in $^{60}$Fe (left) and in $^{59}$Fe (right). The symbols are the SMMC results in the complete $fpg_{9/2}$ shell and the dotted-dashed lines are obtained in the independent-particle model. For $^{60}$Fe we also show the BCS heat capacity (solid line). The two finite discontinuities occur at the proton and neutron pairing transition temperatures. Bottom panels: the number of $n n$, $p p$ and $p n$ $J=0$ pairs as a function of temperature. The `bump' in the heat capacity of $^{60}$Fe is correlated with a rapid reduction in the number $J=0$ neutron pairs.  From Ref.~\cite{LA01}}
\end{figure}

   The top panels of Fig.~\ref{bcs} show the SMMC heat capacity of $^{60}$Fe and $^{59}$Fe (symbols) as a function of temperature. The calculations are done in the $fpg_{9/2}$ model space using the interaction (\ref{nuclear}). We observe that the BCS discontinuity is strongly suppressed but for the even-mass $^{60}$Fe a `bump' remains around the neutron pairing transition temperature. This bump is correlated with the rapid decrease of the number of $J=0$ neutron pairs $\langle \Delta^\dagger \Delta\rangle$ (see bottom left panel of Fig.~\ref{bcs}). However, in the nucleus $^{59}$Fe (containing an odd number of neutrons) we do not observe such a bump in the heat capacity.

\vspace{3 mm}

\begin{figure} [h!]\label{hc-systematics}
  \includegraphics[height=.27\textheight]{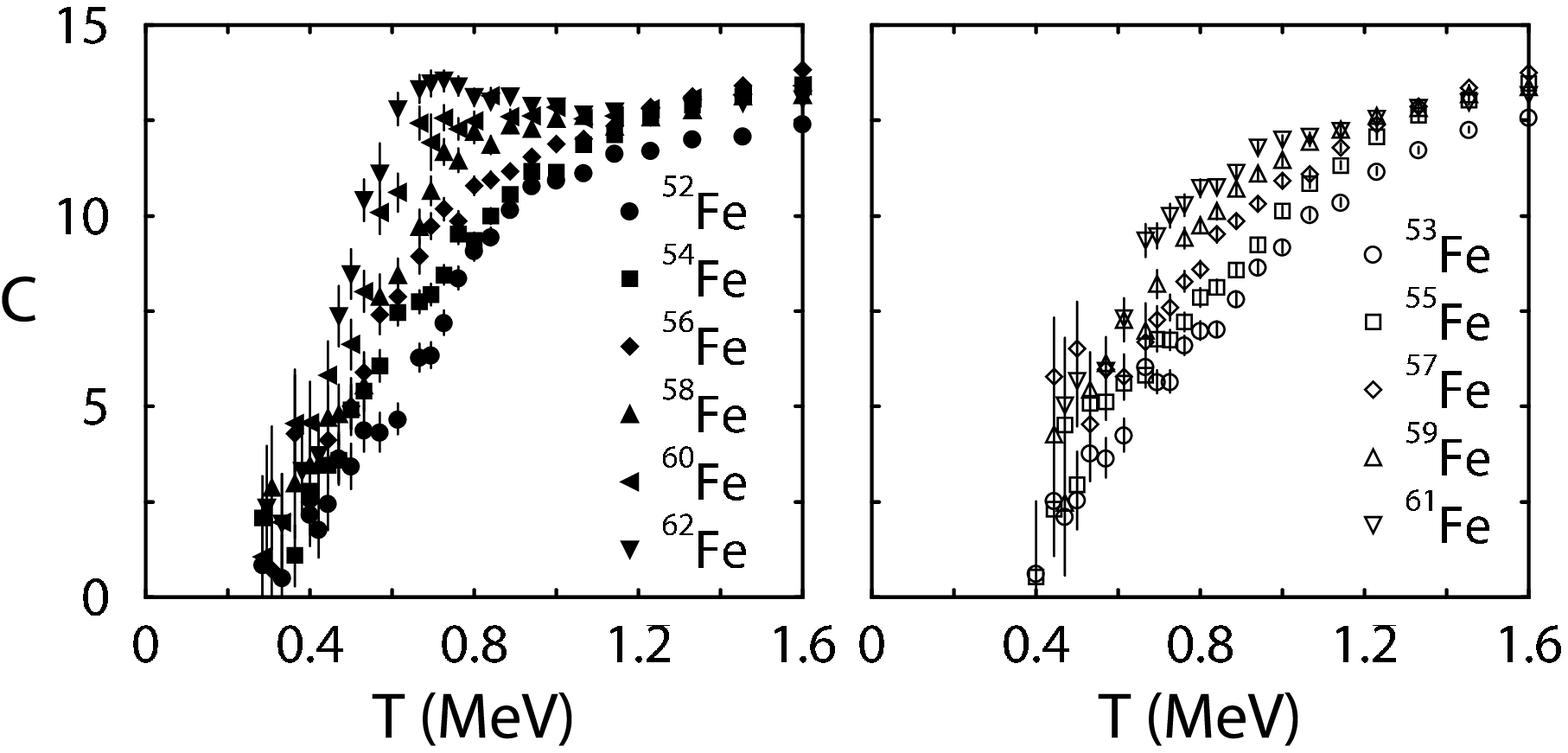}
  \caption{Systematics of the SMMC heat capacity in iron isotopes. Left: heat capacities of even-mass iron isotopes with $A= 52 - 62$. Notice the enhancement of the `bump' with increasing number of neutron pairs. Right: heat capacities of odd-mass iron isotopes with $A=53 - 61$. From Ref.~\cite{LA01}}
\end{figure}

 We expect the above signature of pairing correlations in even-even nuclei to be enhanced upon the addition of neutron pairs. Figure \ref{hc-systematics} shows the SMMC heat capacities (in the $fpg_{9/2}$ model space) for a series of even-mass iron isotopes (left panel) and odd-mass iron isotopes (right panel). The bump observed in the heat capacity of the even-mass iron isotopes is enhanced as the number of neutrons increases, while for the odd-mass iron isotopes we only observe the usual enhancement of the heat capacity with system's size.

\subsubsection{Extending the theory to higher temperatures}\label{extended-theory}

  Since only one major shell is included in the SMMC calculations, the heat capacity saturates at higher temperature. This happens in the vicinity of the bump (see, e.g., in Fig.~\ref{bcs}). To demonstrate that the bump we observe in even-even nuclei is indeed a unique signature of  pairing correlations, it is necessary to increase the model space to include additional shells. We have extended the shell model theory to higher temperatures by combining the fully correlated partition in the truncated model space with the independent-particle partition in the full space (including all bound orbitals and continuum effects) \cite{ABF03}.

  In the absence of interactions, the grand-canonical many-particle partition function of a nucleus with single-particle levels $\epsilon_{nlj}$ and scattering phase shifts $\delta_{lj}(\epsilon)$ is given by
\begin{eqnarray}
\ln Z^{\rm GC}_{sp}  =  \sum_{lj}(2j+1) \left\{\sum_n
\ln [1 + e^{-\beta(\epsilon_{nlj}-\mu)}]  +
\int_0^\infty d\epsilon \delta\rho(\epsilon)\ln[1 +
e^{-\beta(\epsilon-\mu)}]\right\}
\end{eqnarray}
where $\delta\rho(\epsilon) =  \pi^{-1}\sum_{l j} (2j+1)
{d\delta_{lj}/d\epsilon}$ is the continuum contribution to the single-particle level density.

The canonical partition function $Z_N$ at fixed particle-number $A$ can be obtained in the saddle-point approximation
$\ln Z_A \approx \ln Z^{\rm GC} -\beta\mu A -{1\over 2}\ln
\left( 2 \pi  \langle(\Delta N)^2\rangle \right)$,
where $\langle(\Delta A)^2\rangle$ describes the variance of
the particle number fluctuation.

In the presence of interactions we assume
\begin{equation}\label{extended-Z}
 \ln Z_v = \ln Z_{v,tr} + \ln Z_{sp} -\ln Z_{sp,tr} \;.
\end{equation}
Here $Z_v$ is the partition function in the presence of correlations, $Z_{v,tr}$ is the correlated partition in the truncated space, $Z_{sp}$ is the many-particle partition function of the independent-particle model in the full space, and $Z_{sp,tr}$ is the independent-particle model partition in the truncated space. In Eq.~(\ref{extended-Z}) we have subtracted $\ln Z_{sp,tr}$  to avoid a double counting of the truncated degrees of freedom. Eq.~(\ref{extended-Z}) has the correct form at both low temperatures (where $\ln Z_{sp} \approx \ln Z_{sp,tr}$) and at high temperatures (where correlations become less important).

\begin{figure}[h!]\label{hc}
  \includegraphics[height=.5\textheight]{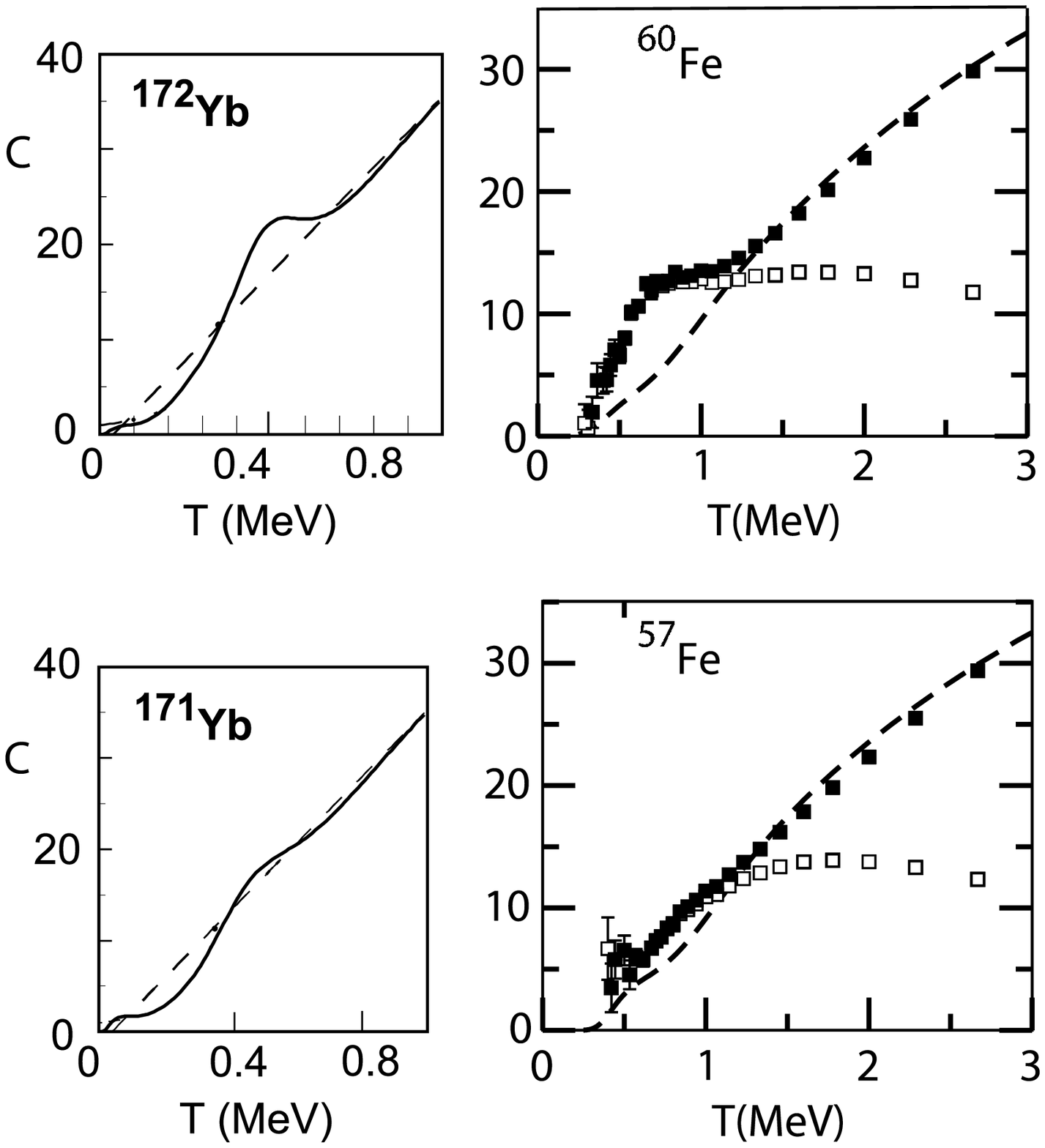}
  \caption{Right: extended heat capacities of $^{60}$Fe (top) and $^{57}$Fe (bottom) versus temperature $T$ \cite{ABF03}. The extended heat capacities (solid squares) are compared with the SMMC heat capacities in the $fpg_{9/2}$ shell (open squares) and with the heat capacities calculated in the independent-particle model (dashed lines). Left: measured heat capacities in $^{172}$Yb  and $^{171}$Yb (solid lines) are compared with  heat capacities in the Fermi gas model (dashed lines) \cite{schiller01}. Even though the experiments are done in a different mass region, the shape of the experimental heat capacities are similar to the calculated heat capacities for both even-even and even-odd nuclei, respectively.}
\end{figure}

\subsubsection{Extended heat capacity}

   We have used the extended theory of Sec.~\ref{extended-theory} to calculate the heat capacity for iron isotopes \cite{ABF03}. The results are shown in the right panels of Fig.~\ref{hc} where the extended heat capacities of $^{60}$Fe and $^{59}$Fe are displayed versus temperature (solid squares).  In contrast with the truncated SMMC calculations (open squares), the extended heat capacity continues to increase monotonically also at higher temperatures. The bump in the heat capacity of the even-even nucleus $^{60}$Fe is now clearly observed in comparison with the heat capacity of the independent-particle model (dashed line).

  Similar $S$-shaped heat capacities were measured in even-even rare-earth nuclei \cite{schiller01}. The experimental heat capacity was determined from the many-particle level density $\rho(E)$ of the corresponding nucleus. This level density was measured in heavy ion reactions and the canonical partition function calculated by a Laplace transform $Z(\beta) = \int d E \rho(E) e^{-\beta E}$. The canonical heat capacity was then determined from $C=dE/dT$ where $E$ is the thermal energy $E=-d\ln Z/ d \beta$.

\section{Conclusion}

  We have discussed several mesoscopic effects in quantum dots, metallic nanoparticles and nuclei. In particular, we have focused on ideas and phenomena in nuclear physics that have found applications in mesoscopic physics and nanoscience.

   Random matrix theory, introduced to explain the statistical features of the neutron resonances in compound nuclei, turned out to be a useful theory for understanding the mesoscopic properties of chaotic ballistic dots. However, whereas in nuclei random matrix theory is used to describe the statistical properties of the excited many-body states, in quantum dots it is used to describe the statistical properties of the single-particle spectrum and wave functions. In almost-isolated dots, electron-electron interactions beyond the classical charging energy have important effects on the mesoscopic fluctuations. A quantitative description can be achieved only when both single-particle chaos and electron-electron interactions are taken into account.

  Studies of metallic nanoparticles have shed light on pairing correlations in finite systems. As the size of the nanoparticle is made smaller, a crossover occurs from the BCS bulk limit of superconductivity to the fluctuation-dominated regime. Of particular interest are pairing correlation effects in this  fluctuation-dominated regime. Nuclei belong to this crossover regime and can also be used to understand the signatures of pairing correlations in finite systems beyond BCS theory. In neutron-rich even-even nuclei we find a `bump' in the heat capacity that is a signature of the pairing transition, while no such signature is found in neighboring nuclei with an odd number of neutrons. These theoretical results are in agreement with recent experiments. A similar effect is expected in the heat capacity of ultrasmall metallic grains. Theoretical studies in metallic nanoparticles and nuclei suggest that pairing correlations in the fluctuation-dominated regime manifest through effects that depend on the particle-number parity.

  Quantum dots, metallic nanoparticles and nuclei are finite-size systems in which the interplay between single-particle behavior and many-body correlations leads to a variety of interesting phenomena.  We have only described some of these phenomena and much remains to be explored in the future.


\begin{theacknowledgments}
   This work was supported in part by the Department of Energy grant No.\ DE-FG-0291-ER-40608. I would like to thank G.F. Bertsch, L. Fang,  L.I. Glazman, M. G\"ok{\c c}eda{\u g}, D. Huertas-Hernando, R.A. Jalabert, A. Kaminski, S. Liu, S. Malhotra, C.M. Marcus, H. Nakada, T. Rupp, A.D. Stone, H.A. Weidenm\"uller and A. Wobst for their collaboration on various parts of the work presented above.
\end{theacknowledgments}

\end{document}